\documentclass[11pt]{article}
\pdfoutput=1  % togliere il commento se le figure sono pdf, lasciarlo se sono eps
\usepackage{graphicx}

\usepackage{appendix}

\usepackage{amssymb}
\usepackage{amsmath}
\usepackage{amsfonts}
\usepackage{upgreek}
\usepackage{latexsym}
\usepackage{slashed,upgreek,amscd,cancel,tensor}
\usepackage{adjustbox}
\usepackage{epsfig}
\usepackage{cite}
\usepackage{hyperref}
\usepackage{amsfonts}
\usepackage{booktabs}
\usepackage{siunitx}
\usepackage{caption}
\usepackage{float}
\usepackage{color}

\numberwithin{equation}{section}% numera le equazioni seconde le sezioni , e.g. 1.15 invece che consecutivamente; anche le appendici, eq. (A.1) etc. Richiede amsmath

\definecolor{MyBlue}{rgb}{0.15,0.15,0.70}

\hypersetup{
colorlinks=true,
citecolor=MyBlue,
linkcolor=MyBlue,
urlcolor=MyBlue
}

\input epsf
\usepackage[top=2cm,bottom=3cm,left=2.6cm,right=2.6cm,asymmetric]{geometry}

\newcommand{\be}{\begin{equation}}
\newcommand{\ee}{\end{equation}}
\newcommand{\beq}{\begin{equation}}
\newcommand{\eeq}{\end{equation}}
\newcommand{\bea}{\begin{eqnarray}}
\newcommand{\eea}{\end{eqnarray}}

\newcommand{\R}{R}

\usepackage[stable]{footmisc}
\def\d{\delta}

\def\Mp{M_*}

\def\s{\sigma}

\def\dkmu2{\delta K_{\mu \nu}\delta K^{\mu \nu}}
\def\pmu2{  \phi_{\mu \nu}\phi^{\mu \nu}}

\newcommand{\aB}{\alphaB}
\newcommand{\MB}{M_{\rm B}}
\newcommand{\MsqK}{M^2_{\rm K}}

\newcommand{\bN}{N_0}

\newcommand{\tg}{\check{g}}

\renewcommand\[{\left[}

\newcommand\ees{\end{eqnarray}}
\newcommand\bees{\begin{eqnarray}}

\newcommand\alphaD{\alpha_{\text{D}}}
\newcommand\alphaC{\alpha_{\text{C}}}
\newcommand\alphaB{\alpha_{\text{B}}}
\newcommand\alphaM{\alpha_{\text{M}}}
\newcommand\alphaK{\alpha_{\text{K}}}
\newcommand\alphaT{\alpha_{\text{T}}}
\newcommand\alphaH{\alpha_{\text{H}}}

\newcommand{\MMM}{{\cal M}}
\newcommand{\SSS}{{\cal S}}

\def\a{\alpha}

\newcommand\h{h}

\newcommand{\omskz}{{\cal E}_1}
\newcommand{\omfkt}{{\cal E}_2}
\newcommand{\omfkz}{{\cal E}_3}

\newcommand{\omtkf}{{\cal E}_4}
\newcommand{\omtkt}{{\cal E}_5}

\newcommand{\tzeta}{{\tilde \zeta}}
\newcommand{\quadac}{{\rm  quad}}
\newcommand{\As}{A}
\newcommand{\Vs}{V}

\newcommand{\VVV}{{\cal V}}

\newcommand{\aone}{{c_{1,0}}}

\newcommand{\athr}{c_{3,0}}
\newcommand{\afour}{c_{4,0}}
\newcommand{\afive}{c_{5,0}}

\newcommand{\asev}{{c_{7,0}}}

\newcommand{\bone}{{c_{1,2}}}
\newcommand{\btwo}{{c_{2,2}}}
\newcommand{\bthr}{{c_{3,2}}}
\newcommand{\bfour}{{c_{4,2}}}

\newcommand{\bsix}{{c_{6,2}}}
\newcommand{\bsev}{{c_{7,2}}}
\newcommand{\beight}{{c_{8,2}}}

\newcommand{\ctwo}{{c_{2,4}}}

\newcommand{\cfour}{{c_{4,4}}}

\newcommand{\dtwo}{{c_{2,6}}}

\newcommand{\dfour}{{c_{4,6}}}

\newcommand{\dels}{\delta\sigma}
\newcommand{\delsD}{\dot{\delta\sigma}}
\newcommand{\dsz}{\dot{\sigma}_0}

\newcommand{\2}{{(2)}}
\newcommand{\3}{{(3)}}
\newcommand{\aK}{\alpha_{\rm K}}
\newcommand{\aL}{\alpha_{\rm L}}
\newcommand{\aH}{\alpha_{\rm H}}
\newcommand{\aT}{\alpha_{\rm T}}
\newcommand{\bun}{\beta_1}
\newcommand{\bdeux}{\beta_2}
\newcommand{\btrois}{\beta_3}
\newcommand{\aq}{a}  % fonctions a_A des DHOST quadratiques 
\newcommand{\bb}{b} % fonctinos b_A des DHOST cubiques
\newcommand{\A }{\As  } %Astar
\newcommand{\V}{\Vs}  % Vstar
\newcommand{\acc}{{\bf a}} % acceleration vector
\newcommand{\Az}{A}
\newcommand{\Bz}{B}
\newcommand{\Cz}{C}
\newcommand{\CI}{{\cal C}_{\rm I}}
\newcommand{\CII}{{\cal C}_{\rm II}}
\newcommand{\CU}{{\cal C}_{\rm U}}

\begin{document}

\begin{center}
\LARGE{\bf Effective Description of \\  Higher-Order Scalar-Tensor Theories}
\\[1cm] 

\large{David  Langlois$^{\rm a}$,  Michele Mancarella$^{\rm b}$, Karim Noui$^{{\rm c},{\rm a}}$, Filippo Vernizzi$^{\rm b}$}
\\[0.5cm]

\small{
\textit{$^{\rm a}$
  APC -- Astroparticule et Cosmologie, 
\\Universit\'e Paris Diderot Paris 7, 75013 Paris, France}}
\vspace{.2cm}

\small{
\textit{$^{\rm b}$ Institut de physique th\' eorique, Universit\'e  Paris Saclay \\ [0.05cm]
CEA, CNRS, 91191 Gif-sur-Yvette, France}}

\vspace{.2cm}

\small{
\textit{$^{\rm c}$ Laboratoire de Math\'ematiques et Physique Th\'eorique, \\ [0.05cm]
Universit\'e Fran\c cois Rabelais, 
\\[0.05cm]
Parc de Grandmont, 37200 Tours, France}}
 %CEA, IPhT, 91191 Gif-sur-Yvette c\'edex, France \\ [0.05cm]
 %\\
%CNRS,  URA-2306, 91191 Gif-sur-Yvette c\'edex, France}}

\end{center}

\vspace{.2cm}

\begin{abstract}
Most existing theories of dark energy and/or modified gravity, involving
a scalar degree of freedom, can be conveniently described within the framework of the Effective Theory of Dark Energy, based on the unitary gauge where the scalar field is uniform.
We extend this effective approach by  allowing  the  Lagrangian in unitary gauge  to depend on the time derivative of the lapse function.  Although this dependence  generically signals the presence of an extra scalar degree of freedom, theories that contain only one propagating scalar degree of freedom, in addition to the usual tensor modes, can be constructed by requiring the initial Lagrangian to be degenerate. Starting from a general quadratic action, we derive the dispersion relations for the linear perturbations around  Minkowski and  a cosmological background. Our analysis directly applies to the recently introduced Degenerate Higher-Order Scalar-Tensor (DHOST)  theories. For these theories, we find that one cannot recover a Poisson-like equation in the static linear regime  except for the subclass that includes the Horndeski and so-called ``beyond Horndeski'' theories. We also discuss Lorentz-breaking models inspired by Horava gravity. 

\end{abstract}
\vspace{.2cm}

\section{Introduction}

The observation of the present cosmological acceleration has spurred the study  of a wide range of theories of dark energy and modified gravity. The number of existing models is now so large that an effective approach 
encompassing as many models as possible is an efficient way to  synthesize
 the various predictions and to confront theoretical models with present and forthcoming data. 
Since many models of dark energy and modified gravity, although not all of them, involve a scalar field in an explicit or implicit way, an effective description based on ADM treatment in the so-called unitary gauge where the scalar field is spatially uniform, is particularly useful and has been actively developed in the last few years.

 Often called Effective Theory of Dark Energy~\cite{Creminelli:2008wc,Gubitosi:2012hu,Bloomfield:2012ff,Gleyzes:2013ooa,Bloomfield:2013efa,Piazza:2013coa,Gleyzes:2014rba}, this approach (see also \cite{Baker:2011jy,Baker:2012zs,Battye:2012eu,Battye:2013ida,Lagos:2016wyv} for  other effective approaches to scalar-tensor theories)
 is inspired by the Effective Field Theory of Inflation~\cite{Creminelli:2006xe,Cheung:2007st} and 
 is based on an action whose building blocks are the lapse $N$, the shift $N^i$ and the spatial metric $h_{ij}$, which all appear  in the ADM metric, 
 \beq
 \label{ADM}
 ds^2=-N^2 dt^2 +\h_{ij} (dx^i+N^i dt)(dx^j+N^jdt)\,.
 \eeq
The shift $N^i$ and the spatial metric $h_{ij}$ appear in the Lagrangian in combinations that behave as 
three-dimensional
tensors under time-dependent spatial diffeomorphisms. 
One such combination is the ``velocity''  of the spatial metric, expressed by the extrinsic curvature tensor $K_{ij}$. Another one is the 3-dimensional Ricci scalar $R$.

The time derivative of the lapse is usually not included in the initial action because the presence of $\dot N$ generically leads to an additional propagating degree of freedom. However, there are special cases where  the action depends on $\dot N$ without leading to an extra degree of freedom.\footnote{When higher time derivative terms can be treated perturbatively below some energy scale, the extra degree of freedom is not excited. Here we consider higher time derivatives at the same level as the other terms.} 
For instance,  starting from an action whose ADM form in the unitary gauge does not contain any $\dot N$ and   making a conformal transformation of the metric that depends on the scalar field gradient leads to an action with an $\dot N$ dependence.
 In that case, the presence of $\dot N$ terms is not problematic because there is a degeneracy in the kinetic terms, which prevents the existence of a ghost-like degree of freedom 
(see \cite{Gleyzes:2014rba} and \cite{Domenech:2015tca}). 
 
 In the present work, we consider systematically Lagrangians quadratic in linear perturbations that contain time (and space) derivatives of  $\delta N$, such as to include all possible terms containing at most
two (space or time) derivatives.\footnote{For instance, since $\R$ contains two spatial derivatives, we do not include a term such as $\R\,   \delta \dot N$, which depends on three derivatives.} The corresponding quadratic 
action, in an expansion around the flat Friedmann-Lema\^itre-Robertson-Walker (FLRW) metric $ds^2 = - dt^2 + a^2(t) d\bf x^2$, can be written in  the form 
\be
\begin{split}
\label{SBAction0}
& S^\quadac = \int d^3x \,  dt \,  a^3  \frac{M^2}2\bigg\{ \delta K_{ij }\delta K^{ij}- \left(1+\frac23\aL\right)\delta K^2  +(1+\alphaT) \bigg( \R \frac{\delta \sqrt{h}}{a^3} + \delta_2 R \bigg)\\
&  + H^2\alphaK \delta N^2+4 H \alphaB \delta K \delta N+ ({1+\alphaH}) \R  \delta N   +  4 \bun  \delta K  {\delta \dot N }   + \bdeux  {\delta \dot N}^2 +  \frac{\btrois}{a^2}(\partial_i \delta N )^2   
\bigg\} \; ,
\end{split}
\ee
where $H \equiv \dot a/a$ is the Hubble rate and $\delta_2R$ stands for the second order term in the perturbative expansion of $R$. Although the spatially diff-invariant combination denoting the ``velocity'' of the lapse is $\dot N - N^i \partial_i N$, the action above contains only $\delta \dot N$, to which the full combination reduces at linear order.
%Note that  the spatially diff-invariant  combination $\dot N - N^i \partial_i N$  reduces to $\delta \dot N$ at linear order, which is why the full combination does not appear in the quadratic perturbative action (\ref{SBAction0}).

The above quadratic action extends the one derived in \cite{Gleyzes:2013ooa} and written in terms of the dimensionless time-dependent functions $\alpha_A$ (introduced in \cite{Bellini:2014fua} and \cite{Gleyzes:2014qga}) in \cite{Gleyzes:2014rba}, with the addition of  four new  functions of time: 
the parameter $\aL$,
%which is nonzero when the relation between the $K_{ij}K^{ij}$ and $K^2$ terms is detuned from the general relativity combination,
and the three parameters  $\beta_A$ that characterize the terms containing (time or space) derivatives of $\delta N$. These parameters can be given the following interpretation:
\begin{itemize}
\item $\aL$ corresponds to a detuning of the extrinsic curvature terms. 
When $\aL=0$ one recovers the combination $K_{ij}K^{ij}-K^2$, which is part of the four dimensional Ricci scalar (via the Gauss-Codazzi identity).
%
%Its presence is reminiscent of the fact that the two terms $K_{ij}K^{ij}$ and $K^2$ are \emph{separately} invariant under space diffs, while for $\aL=0$ one recovers the combination $K_{ij}K^{ij}-K^2$ which comes from the four dimensional Ricci scalar. 
This detuning appears in theories that already in their original formulation assume a preferred time slicing, such as Horava gravity~\cite{Horava:2009uw} and its extensions~\cite{Blas:2009qj,Blas:2009yd,Blas:2010hb,Gao:2014soa}.
\item $\bun$ is analogous to the kinetic braiding $\alphaB$ for the additional degree of freedom present in higher-order theories. % It contributes to its kinetic energy via backreaction through gravity. When this function is non vanishing, the propagating scalar mode is a mixing of the metric perturbations and the lapse perturbations.
\item $\bdeux$, similarly, is the analogue of the kineticity $\alphaK$. 
\item $\btrois$ is associated to the gradient energy of the additional degree of freedom. This comes from the acceleration of the unit vector normal to the uniform scalar field hypersurfaces, which in unitary gauge is given by $a_i=\partial_i N/N$.
\end{itemize}
%In section~\ref{sec:3} we will argue that the above properties become manifest in the dispertion relation of the propagating scalar modes over a Minkowski background.

As will be shown explicitly in Sec.~\ref{section_DHOST}, one can obtain an action of the form~\eqref{SBAction0} by starting from a covariant scalar-tensor  action and  choosing a slicing where the scalar field $\phi$ depends only on time. Whereas usual scalar-tensor Lagrangians, which depend only on $\phi$ and its first order gradient $\nabla_\mu\phi$, lead to 
%DL
effective perturbative actions 
where  only $\aK$ can be nonzero, scalar-tensor Lagrangians that depend as well on second-order derivatives $\nabla_\mu\!\nabla_\nu\phi$, lead to a much richer phenomenology. Allowing for higher-order derivatives in the Lagrangian is potentially dangerous as, in general, this yields higher-order equations of motion requiring extra initial conditions, thus signalling the presence of  an extra scalar degree of freedom associated with instabilities. 

However, it is possible to find higher-order scalar-tensor theories that contain a single scalar degree of freedom (in addition to the tensor modes associated with gravity) by imposing some appropriate restrictions on the initial Lagrangian. For instance, requiring that the associated Euler-Lagrange equations
are second order leads to Horndeski theories \cite{Horndeski:1974wa}, associated with nonzero $\aK$, $\aB$ and  $\aT$ (while $\aL$, $\aH$ and the $\beta_{A}$  vanish). Overcoming the prejudice that second order  equations of motion were necessary to get only one propagating scalar mode, the introduction of a larger class of models,   often called ``beyond Horndeski'' theories,  showed that the absence of an extra scalar mode is compatible with third order
Euler-Lagrange equations\cite{Gleyzes:2014dya,Gleyzes:2014qga}.\footnote{See also \cite{Zumalacarregui:2013pma} for an earlier example based on the disformal transformation of the Einstein-Hilbert Lagrangian.} These ``beyond Horndeski'' theories give a nonzero $\aH$, but $\aL$ and the $\beta_A$ are still vanishing.

In \cite{Langlois:2015cwa,Langlois:2015skt}, it was realized that all higher-order scalar-tensor theories that contain a single scalar mode can be understood as degenerate theories, dubbed Degenerate Higher-Order Scalar-Tensor (DHOST)
theories.\footnote{The  theories discovered in \cite{Langlois:2015cwa} were also named Extended Scalar-Tensor (EST) theories in \cite{Crisostomi:2016czh}.  We prefer to use the more specific  terminology of DHOST theories, introduced in \cite{Achour:2016rkg}.}
Here, degenerate means that the Hessian matrix obtained by taking the second derivatives of the Lagrangian with respect to velocities\footnote{More precisely, as explained in detail in \cite{Langlois:2015cwa}, one first introduces  an auxiliary variable that includes the time derivative of the scalar field $\dot\phi$, as well as the lapse and the shift, so that all second-order time derivatives $\ddot \phi$ are absorbed by the velocity of this auxiliary variable and there are no longer time derivatives of the lapse and of the shift. Thus, the kinetic part of the resulting Lagrangian does not depend anymore on an acceleration but  just  on velocities.} 
is a degenerate matrix (see~\cite{Motohashi:2016ftl,Klein:2016aiq,Crisostomi:2017aim} for recent considerations on the notion of degeneracy). All DHOST theories up to quadratic order, i.e. whose Lagrangian depends quadratically on $\nabla_\mu\!\nabla_\nu\phi$, were identified in \cite{Langlois:2015cwa}.  The systematic classification of DHOST theories up to cubic order was recently completed in \cite{BenAchour:2016fzp}.

Horndeski  and ``beyond Horndeski'' theories are included in the class of  DHOST theories.  In fact, they belong to the same subclass of DHOST theories and can be related to each other via disformal transformations \cite{Gleyzes:2014qga,Crisostomi:2016tcp,Crisostomi:2016czh,Achour:2016rkg,Ezquiaga:2017ner}. But, along this special subclass that contains Horndeski  and ``beyond Horndeski'' theories, DHOST theories include many other subclasses of theories: six other subclasses in the purely quadratic case, eight in the purely cubic case and 24 other subclasses for theories with both quadratic and cubic terms. 

The effective description of dark energy models is a powerful tool to confront models with present and future observations, see for instance~\cite{Piazza:2013pua,Ade:2015rim,Perenon:2015sla,Frusciante:2015maa,Gleyzes:2015rua,Bellini:2015xja,Hu:2016zrh,Salvatelli:2016mgy,Leung:2016xli,Renk:2016olm,Bettoni:2016mij,DAmico:2016ntq,Alonso:2016suf,Bellomo:2016xhl}. It is also an efficient way to classify the phenomenology of various theories~\cite{Pogosian:2016pwr,Perenon:2016blf}. So far, the effective approach has mainly been used for Horndeski and beyond Horndeski theories, although it has also been extended to include models such as Horava gravity\cite{Kase:2014cwa,Frusciante:2015maa,Frusciante:2016xoj,DeFelice:2016ucp}.
The purpose of this work is to generalize this effective approach in order to include DHOST theories.

The layout of this paper and our main results 
can be summarized as follows.  In the next section, we briefly
present the DHOST theories (up to cubic order) 
and derive their Lagrangian in the unitary gauge.
The degeneracy of DHOST theories implies that the parameters in action (\ref{SBAction0}) cannot be arbitrary but must satisfy some consistency relations. We find that there are two such sets 
of degeneracy conditions, 
given in Sec.~\ref{sec:2.2}, 
which we name $\CI$ and $\CII$:
  they  relate the parameters $\bun$, $\bdeux$, $\btrois$, with $\aL$, $\aH$ and $\aT$.   The first set of conditions is characterized by $\aL=0$, while $\bun$ remains arbitrary. By contrast, in the second set, 
$\aL$ is arbitrary while all $\beta$'s are fully determined in terms of $\aH$, $\aT$ and $\aL$. This implies that all the DHOST theories we investigate can be regrouped into three main families: those satisfying only $\CI$, those verifying only $\CII$, and finally the theories for which both sets of conditions $\CI$ and $\CII$ are valid. 

In Sec.~\ref{sec:2.3}, we  also study
  how the action \eqref{SBAction0} transforms under the most general conformal-disformal transformation, allowing the conformal and disformal factors to depend on the scalar field, as well as on  $X\equiv \nabla_\mu \phi \nabla^\mu\phi$. 
After the transformation, the action takes the same form as \eqref{SBAction0}, 
with its parameters  related 
to those of eq.~\eqref{SBAction0} by the transformations given in eq.~\eqref{alphatilde}.
In general, all the parameters, except $\aL$, are modified  but we show that both sets of conditions $\CI$ and $\CII$ are preserved under these transformations. 
 The two sets of conditions $\CI$ and  $\CII$ share a  common condition, which  implies that only one scalar mode appears {\it in the unitary gauge},  but this condition is not enough to guarantee that this remains
true in an arbitrary gauge.

The family of DHOST theories that satisfy $\CI$ but not $\CII$ coincides with theories that are related to Horndeski via (conformal-) disformal transformations. 
 For all the other theories, i.e.~those satisfying $\CII$, we find that the effective 
 Newton constant in  the analog of the Poisson equation becomes infinite,  as a  direct consequence of one of the conditions in  $\CII$. 
 Therefore, one cannot recover a Poisson-like equation in the static linear regime for these theories, 
 in contrast with theories   verifying only $\CI$,  where $\bun$ is unconstrained. 
 If this peculiar behaviour persists at the nonlinear level, this would indicate that only theories  that are related to Horndeski via conformal or disformal transformations are phenomenologically viable.
 
 We examine the dispersion relation for scalar modes around Minkowski in Sec.~\ref{sec:3} and in a cosmological background 
 in Sec.~\ref{sec:cosmo}. 
 In both cases, we observe that the  dispersion relation $\omega^2 = \omega^2(k^2)$ is in general a  rational 
 function of $k^2$.
This drastically simplifies to a linear dispersion relation $\omega^2 = c_s^2 k^2$   when the degeneracy conditions $\CI$ or  $\CII$ are satisfied. 
  In the cosmological context, we also derive the quadratic action 
  for the curvature perturbation on uniform field hypersurfaces $\zeta$ and show that it is conserved on super-Hubble scales.
Then, in Sec.~\ref{sec:5} we discuss
 two classes of Lorentz-breaking theories that have been introduced in the literature. Finally, we present some conclusions in the final section. We have also added several appendices, where more technical details are provided.

%%%%%%%%%%%%%%%%%%%%%%%%%%
\section{DHOST theories}
%%%%%%%%%%%%%%%%%%%%%%%%%%
\label{section_DHOST}

In this section, we present  a large class  of scalar-tensor theories whose action, which depends on  a metric $g_{\mu\nu}$ and a scalar field $\phi$, leads to a quadratic action   of the form (\ref{SBAction0}) when written in the unitary gauge.  More precisely, we assume that the Lagrangian depends not only on $\phi$ and its gradient $\phi_\mu\equiv\nabla_\mu\phi$ as  usual,  but also on its second derivatives $\phi_{\mu\nu}\equiv\nabla_\mu\!\nabla_\nu\phi$.

\subsection{Covariant action}
 Allowing for a dependence on $\phi_{\mu\nu}$ up to cubic order, we consider an action  of the form
 \bea
\label{action}
S[g,\phi] &=& \int d^4 x \, \sqrt{- g }
\left[ P(X,\phi) + Q(X,\phi) \Box \phi
+
f_2(X,\phi) \,  {}^{(4)}\!R+ C_\2^{\mu\nu\rho\sigma} \,  \phi_{\mu\nu} \, \phi_{\rho\sigma}
\right.
\cr
&&
\left. \qquad \qquad\qquad + f_3(X, \phi) \, {}^{(4)}G_{\mu\nu} \phi^{\mu\nu}  +  
C_\3^{\mu\nu\rho\sigma\alpha\beta} \, \phi_{\mu\nu} \, \phi_{\rho\sigma} \, \phi_{\alpha \beta} \right]  \;,
\eea
where the functions $f_2$  and $f_3$ depend only on the scalars $\phi$ and $X \equiv \phi_\mu \phi^\mu$;  ${}^{(4)}\!R$ and ${}^{(4)}G_{\mu\nu}$ denote, respectively, the usual Ricci scalar and Einstein tensor associated with the metric $g_{\mu\nu}$. 

The tensors 
$C_\2$ and $C_\3$ are the most
general tensors constructed from the metric $g_{\mu\nu}$ and the first derivative of the scalar field 
$\phi_\mu$.
It is easy to see that the quadratic terms can be written as
\beq
\label{C2}
 C_\2^{\mu\nu\rho\sigma} \,  \phi_{\mu\nu} \, \phi_{\rho\sigma} =\sum_{A=1}^{5}a_A(X,\phi)\,   L^\2_ A\,,
\eeq
with 
\be
\label{QuadraticL}
\begin{split}
& L^\2_1 = \phi_{\mu \nu} \phi^{\mu \nu} \,, \qquad
L^\2_2 =(\Box \phi)^2 \,, \qquad
L_3^\2 = (\Box \phi) \phi^{\mu} \phi_{\mu \nu} \phi^{\nu} \,,  \\
& L^\2_4 =\phi^{\mu} \phi_{\mu \rho} \phi^{\rho \nu} \phi_{\nu} \,, \qquad
L^\2_5= (\phi^{\mu} \phi_{\mu \nu} \phi^{\nu})^2\,.
\end{split}
\ee
Similarly, the cubic terms can be written as
\beq
\label{C3}
C_\3^{\mu\nu\rho\sigma\alpha\beta} \, \phi_{\mu\nu} \, \phi_{\rho\sigma} \, \phi_{\alpha \beta} = \sum_{A=1}^{10} b_A(X,\phi)\,  L^\3_A \,,
\eeq
where 
\be
\label{CubicL}
\begin{split}
& L^\3_1=  (\Box \phi)^3  \,, \quad
L^\3_2 =  (\Box \phi)\, \phi_{\mu \nu} \phi^{\mu \nu} \,, \quad
L^\3_3= \phi_{\mu \nu}\phi^{\nu \rho} \phi^{\mu}_{\rho} \,,   \\
& L^\3_4= \left(\Box \phi\right)^2 \phi_{\mu} \phi^{\mu \nu} \phi_{\nu} \,, \quad
L^\3_5 =  \Box \phi\, \phi_{\mu}  \phi^{\mu \nu} \phi_{\nu \rho} \phi^{\rho} \,, \quad
L^\3_6 = \phi_{\mu \nu} \phi^{\mu \nu} \phi_{\rho} \phi^{\rho \s} \phi_{\s} \,,   \\
& L^\3_7 = \phi_{\mu} \phi^{\mu \nu} \phi_{\nu \rho} \phi^{\rho \s} \phi_{\s} \,, \quad
L^\3_8 = \phi_{\mu}  \phi^{\mu \nu} \phi_{\nu \rho} \phi^{\rho}\, \phi_{\s} \phi^{\s \lambda} \phi_{\lambda} \,,   \\
& L^\3_9 = \Box \phi \left(\phi_{\mu} \phi^{\mu \nu} \phi_{\nu}\right)^2  \,, \quad
L^\3_{10} = \left(\phi_{\mu} \phi^{\mu \nu} \phi_{\nu}\right)^3 \,.
\end{split}
\ee

In general, theories with an action of the form (\ref{action}), which  depends on 
second-order derivatives of $\phi$, contain two tensor modes and two scalar modes, one of which is associated with a so-called  Ostrogradsky instability \cite{Ostrogradsky:1850fid,Woodard:2015zca}. 
 However, it is possible to choose special  functions $\aq_A$ and $\bb_A$  in the terms of the Lagrangian (\ref{C2}) and (\ref{C3})  so that  the corresponding theory   is degenerate and  contains at most one propagating scalar mode.
 This class of theories, also known as  DHOST theories, has originally been identified  at  quadratic order
 %DL
 in $\phi_{\mu\nu}$
  (i.e.~with the functions $f_2$ and $\aq_A$ only)  
  in \cite{Langlois:2015cwa} and further studied in \cite{Langlois:2015skt,Crisostomi:2016czh,Achour:2016rkg,deRham:2016wji} (see also~\cite{Ezquiaga:2016nqo} for an approach to scalar-tensor theories based on differential forms). 

The identification of DHOST theories has recently been extended up to cubic order, i.e.~by including the second line of (\ref{action}), in \cite{BenAchour:2016fzp} and the interested reader will find the full classification there
 (see Table \ref{table_DHOST} 
 %in App.~\ref{app_degeneracy} 
 %DL
 for a short summary).  
The DHOST theories include all Horndeski theories but also new theories
%DL
that
 lead to higher-order 
Euler-Lagrange equations  
even if no extra scalar mode propagates. In summary, there exist seven  classes of purely quadratic theories (four classes with $f_2\neq 0$ and three classes with $f_2=0$) and  nine classes of purely cubic theories (two with $f_3\neq 0$ and seven with $f_3=0$). These quadratic and cubic classes can be combined to yield hybrid theories, involving both quadratic and cubic terms, but all combinations are not possible: only 25 combinations (out of 63)  lead to degenerate theories, 
often with extra conditions on the functions $a_A$ and $b_A$ in the Lagrangian (see \cite{BenAchour:2016fzp} for  details).

\subsection{(3+1) decomposition in the unitary gauge}
We now wish to reexpress the action (\ref{action}) in ADM form in the unitary gauge. For simplicity, we discuss here only the quadratic case. More details about the calculations and their extension to  cubic theories are given in App.~\ref{app_unitary}.

In order to write the $(3+1)$ decomposition of the  action 
(\ref{action}),  it is convenient to use the notation of \cite{Langlois:2015cwa,Langlois:2015skt}
and   introduce the 
auxilary variables\footnote{We have slightly changed the notation by using $\A$ and $\V$ instead of $\A_*$ and $\V_*$.} 
\beq
\A\equiv\frac1N{\cal D}_t\phi\equiv\frac1N (\dot\phi-N^i \partial_i\phi)\,, \qquad
\V\equiv \frac1N{\cal D}_t \A  \,.
\eeq
The action (\ref{action}) can then be expressed in terms of $\V$, corresponding to the velocity of $\A$, and of the extrinsic curvature tensor, 
\beq
K_{ij}\equiv \frac{1}{2N}\left(\dot\h_{ij}-D_i N_j -D_jN_i\right)\,,
\eeq
%DL
where $D_i$ denotes the covariant derivative associated with the spatial metric $h_{ij}$.
The full expression for 
%DL
the action in 
an arbitrary gauge can be found in \cite{Langlois:2015skt}. 

Here, we  restrict our derivation to the so-called unitary gauge, where the scalar field is uniform, i.e.
such that 
\beq
\label{UGphi}
\partial_i\phi=0\qquad ({\rm unitary\  gauge})\,.
\eeq
In the unitary gauge, the quantities $\A$ and $\V$ defined above reduce to
\beq
\A =\frac{\dot\phi}{N}\,, \qquad 
\V\equiv \frac1N\left(\dot\A+A\, \frac{N^i\partial_i N}{N}\right) \qquad ({\rm unitary\  gauge})\,.
\eeq

Ignoring the $P$ and $Q$ terms of the Lagrangian, which do not play any role for the degeneracy, 
we can  compute the ADM form of the elementary quadratic and cubic
Lagrangians \eqref{QuadraticL} and \eqref{CubicL} in the unitary gauge. Their expressions
are given in App.~\ref{app_unitary}. One can also obtain the analogous expression for  the terms $f_2 {}^{(4)}\!R$ and $f_3  {}^{(4)}\!G_{\mu\nu}\phi^{\mu\nu}$ by using the Horndeski Lagrangians, as explained in the appendix. 

In the quadratic case, we find that the total  ADM action in the unitary gauge is given by  
\beq
\label{action_quad_unitary}
S = \int  \, d^3x  \,  dt \, N   \sqrt{\h} \,  {\cal L}\,,
\eeq
with 
\be 
\label{LL2}
\begin{split}
{\cal L} = & \ f_2 R - 2 f_{2\phi} \A\,K + (f_2 +\aq_1 \A^2) K_{ij} K^{ij} - (f_2 -\aq_2 \A^2) K^2  \\
& + \left[\aq_1 +\aq_2 -(\aq_3 + \aq_4) \A^2 + \aq_5 \A^4\right] \Vs^2  + \A(4 f_{2X} +2 \aq_2 -\aq_3 \A^2 )  K \V \\
& +\left[  4 f_{2X} \A^2  -(2 \aq_1 -\aq_4 \A^2)\A^2 \right] \frac{\partial_i N\partial^iN}{N^2}\;.
\end{split}
\ee
The full expression including the cubic terms is much more involved and is given in (\ref{LL1}) of App.~\ref{app_unitary}.

We can further simplify  the above expressions by assuming that the scalar field is proportional to the time coordinate $t$,
\beq
\label{muphi}
\phi=\mu^2 t\,,
\eeq
where $\mu$ is some mass scale,
so that 
\beq
\A=\frac{\mu^2}{N}\,, \qquad \V=-\frac{\mu^2}{N^3}{\cal D}_tN\,.
\eeq
In this case, the dynamical quantities are the lapse $N$ and the spatial metric. 

Upon  expanding  the above action (\ref{action_quad_unitary}) 
%DL
around a cosmological background 
up to quadratic order in perturbations, one obtains an expression of the form (\ref{SBAction0}), with 
\beq
\label{effective}
\begin{split}
\frac{M^2}{2}=&\, f_2-\aq_1 X\,, \qquad \frac{M^2}{2}(1+\a_T)= f_2\,, \qquad  \frac{M^2}{2}(1+\a_H)=f_2-2X f_{2X} \,,
\\
\frac{M^2}{2}\left(1+\frac23\aL\right)=&\, f_2+\aq_2 X\,,\qquad \frac{M^2}{2} \bdeux=-X \left(\aq_1+\aq_2+(\aq_3+\aq_4) X+\aq_5X^2\right)\,,
\\
2 M^2\bun=&\, X (4f_{2X}+2\aq_2+\aq_3X) \,, \quad \frac{M^2}{2} \beta_3=-X(4f_{2X}-2\aq_1-\aq_4 X)\,,
\end{split}
\eeq
where the right-hand side quantities are evaluated on the homogeneous and isotropic background (so that $X=-\mu^4$). Let us stress that the coefficients $\bun$ and $\bdeux$ correspond to the terms in front of $K \V$ and $\V^2$, respectively, in the unitary action. This means that all the $\dot N$ terms disappear when $\bun=\bdeux=0$. 

The first two relations of (\ref{effective})  can be used to express $M^2$ and $\a_T$ in terms of $f_2$ and $\aq_1$. Substituting into the other relations, one easily gets the other parameters, $\aH$, $\aL$, $\bun$, $\bdeux$ and $\btrois$, in terms of $f_2$, $f_{2X}$ and $\aq_A$ evaluated on the background. In App.~\ref{app_unitary}, we also give the full expressions of the effective parameters when the action also contains the cubic terms, thus in terms of $f_3$ and the $b_A$.

Let us briefly discuss the values of these parameters for the quadratic DHOST theories. 
The classes Ib, IIb and IIIc  are pathological, as  noted in \cite{Langlois:2015skt}, 
because they do not contain propagating gravitons. Indeed, one sees  immediately that   if $\aq_1=f_2/X$, which is the case for these three classes,   the coefficient of the kinetic term for the gravitons $K_{ij} K^{ij}$ disappears (since $X=-\A^2$ in the unitary gauge), i.e. $M=0$ and the theory does not contain tensor degrees of freedom. 

The remaining theories, IIIa and IIIb, are also problematic, as pointed out in \cite{deRham:2016wji}. Indeed, $f=0$ implies that there is no gradient term for the gravitons since the spatial curvature $R$ disappears. This means that the propagation speed for gravitational waves is zero, or equivalently, $\aT=-1$. Note that the classes IIIa  and IIIb  also verify the  property $\aH=-1$.

From a phenomenological point of view, the  classes Ia and IIa therefore appear to be the most interesting.

\subsection{Degeneracy conditions for the effective parameters}
\label{sec:2.2}
Among theories  of the form (\ref{action}), DHOST theories play a very special role as  their  Lagrangian is degenerate, 
which implies that they contain at most three propagating degrees of freedom, i.e. two tensor modes and one scalar degree of freedom. 

Interestingly, 
the 
%DL
fully nonlinear 
degeneracy conditions boil down to two sets of very simple conditions for the effective parameters appearing in the quadratic
%DL
perturbative
 action. 
Depending on the DHOST theory under consideration, we find that the effective parameters  satisfy either
\beq
\label{Ia}
\CI:\qquad \aL=0\,, \qquad \bdeux=-6\bun^2\,,\qquad   \btrois=-2\bun\left[2(1+\a_H)+\bun (1+\a_T)\right]\,,
\eeq
or the set of conditions
\beq
\label{IIa}
\CII:\qquad \bun=- (1+\aL)\frac{1+\a_H}{1+\a_T}\,, \quad \bdeux=-6(1+\aL) \frac{(1+\a_H)^2 }{(1+\a_T)^2}\,,\quad \beta_3=2\frac{(1+\a_H)^2}{1+\a_T}\,,
\eeq
where we have assumed that $\aT\neq -1$ in the latter case (otherwise\footnote{A model  for which $\aT=-1$ is very peculiar since the speed of gravitational waves vanishes.}  one should use a regular version of the conditions obtained by multiplying both sides of the equalities by the denominator of the right hand side).
It is immediate  to see that both sets of conditions share the common condition
\beq
\label{C_U}
\CU:\qquad (1+\aL)\bdeux=-6\bun^2\,,
\eeq
which plays a special role in the unitary gauge, as we will see later. In the second set of conditions, $\CII$, the three parameters $\beta_A$ are completely determined by the parameters $\aL$, $\aH$ and $\aT$. By contrast, in the set of conditions $\CI$, $\bun$ remains independent of the $\a_A$. Note that a theory that satisfies $\aL=0$ and the conditions $\CII$ automatically verifies $\CI$.
One can also recover directly the conditions $\CI$ and $\CII$ by rewriting the three degeneracy conditions involving $f_2$, $f_{2X}$ and the five functions $\aq_A$, derived in \cite{Langlois:2015cwa}, in terms of the seven parameters  $M^2$, $\aL$, $\aH$, $\aT$ and $\beta_A$, as we show in App.~\ref{app_degeneracy}.

The degeneracy conditions satisfied by each DHOST subclass are indicated in  Table \ref{table_DHOST}.
%of App.~\ref{app_degeneracy}.
Among purely quadratic theories, the subclass\footnote{For  quadratic theories, several names have been introduced in 
%DL
previous works. 
Here we use the names introduced in~\cite{Achour:2016rkg} for quadratic theories. Other names have been used in~\cite{BenAchour:2016fzp} and are reported in Table \ref{table_DHOST}.} Ia satisfies the conditions $\CI$, while the subclass IIa satisfies the conditions $\CII$. 
As mentioned earlier, the effective coefficients cannot be defined for the theories Ib, IIb and IIIc for which $M^2=0$. Theories IIIa and IIIb satisfy $\aT=\aH=-1$ and 
%therefore  
verify the regular version of conditions $\CII$.

The situation with cubic DHOST theories is subtler. The reason is that there are more than three degeneracy conditions for the 11 functions that parametrize the space of cubic scalar-tensor theories, as shown in  \cite{BenAchour:2016fzp}. However,  for linear perturbations about a cosmological background, these degeneracy conditions simply ``project" 
%DL
onto   $\CI$ or $\CII$.   If we pushed the effective description of a cubic DHOST theory up to higher order, we  would expect  the emergence of new degeneracy conditions, which would be reminiscent of the full degeneracy conditions obtained in the complete theory.

For the purely cubic theories, 
one can discard six subclasses out of nine, because they lead to $M^2=0$.
Among the remaining three subclasses,  the subclass ${}^3$N-I, which includes the quintic Horndeski Lagrangian, satisfies the conditions $\CI$, while the other two, ${}^3$M-I and ${}^3$M-II,   obey the conditions $\CII$
%DL
(see Table \ref{table_DHOST}).
%of App.~\ref{app_degeneracy}).

Finally, let us discuss the combinations of quadratic and cubic theories. As shown in \cite{BenAchour:2016fzp}, there exist 25 subclasses of degenerate theories. Only one subclass, Ia\, \& ${}^3$N-I, satisfies the conditions $\CI$ only: this subclass contains the full Horndeski theory as well as the beyond-Horndeski extensions. 
Leaving aside seven subclasses
%DL
%\footnote{\bf They correspond to the allowed combinations of quartic theories with $M^2=0$ and cubic theories with $M^2=0$, namely ${}^2$N-II\ \&  ${}^3$M-VII, ${}^2$N-IV\ \& ${}^3$M-V, ${}^2$M-III\ \& ${}^3$M-V, ${}^2$M-III\ \& ${}^3$M-VI and ${}^2$M-III\ \& ${}^3$M-VII.} 
for which $M^2=0$, we are left with 17 subclasses that satisfy the conditions $\CII$. Among these, 
one subclass also satisfies $\aL=0$ (and therefore the conditions $\CI$ too):  Ia\, \& ${}^3$M-III. 
The other subclasses satisfy only the conditions $\CII$.

It was shown in \cite{Achour:2016rkg} that all subclasses of quadratic DHOST theories are stable with respect to conformal-disformal transformations, by which we mean that any theory is mapped into another theory belonging to the same subclass. One can conjecture that this should remain true for the cubic DHOST theories, although it has been checked only for the subclass ${}^3$N-I containing Horndeski. 
Given these considerations, it is instructive to explore how the effective coefficients are transformed under conformal-disformal transformations.

\subsection{Disformal transformations}
\label{sec:2.3}
\newcommand\alphaX{\alpha_{\text{X}}}
\newcommand\alphaY{\alpha_{\text{Y}}}
Let us  consider  general (conformal-)disformal transformations, which define a new metric  by using the scalar field, according to the expression \cite{Bekenstein:1992pj}
\beq
\label{disf_cov}
 \tilde{g}_{\mu \nu} = C(\phi,X) g_{\mu\nu}+ D(\phi,X) \phi_\mu \phi_\nu.
 \eeq
 As shown explicitly in   \cite{Achour:2016rkg} 
 for quadratic theories, any DHOST theory can be mapped into another DHOST theory via this transformation. More precisely, if we start from a theory defined by the action $\tilde{S}[\tilde{g},\phi]$, one can define a new theory as 
 \beq
 S[g_{\mu\nu},\phi]:=\tilde{S}[\tilde{g}_{\mu\nu}= C g_{\mu\nu}+ D \phi_\mu\phi_\nu, \phi] \,.
 \eeq
 The explicit transformation of the functions $f_2$ and $\aq_A$ can be found in \cite{BenAchour:2016fzp}, where it is also shown that all subclasses of quadratic DHOST theories are stable under disformal transformations.  If the disformal transformation is invertible, 
 i.e.~it satisfies the condition~\cite{Bekenstein:1992pj}
 \be
 \label{invertible}
C - X C_X -  X^2 D_X\neq 0    \;,
 \ee
 then two disformally related theories are equivalent,  provided matter is ignored.
 However, including matter and assuming  that it is  minimally coupled to the metric that appears in  the two disformally related DHOST actions, one gets two physically distinct theories.  
 
In order to compute the transformation of the effective parameters of the quadratic action, it is convenient to introduce the dimensionless time-dependent parameters 
\be
\label{defalphas2}
\begin{split}
&\alphaC \equiv \frac{\dot \phi}{2 H C } \frac{\partial C}{\partial \phi} \; , \quad \alphaY \equiv - \frac{X}{C} \frac{\partial C}{\partial X} \, , \quad  \alphaD \equiv - \frac{D}{ D+ C/X}\, , \quad \alphaX \equiv - \frac{X^2}{ C } \frac{\partial D}{\partial X}\; ,
 \end{split}
\ee
where the right-hand sides are evaluated on the background.
These four dimensionless functions characterize how the quadratic action (\ref{SBAction0}) transforms under the transformation \eqref{disf_cov}. 
The functions $\alphaC$ and $\alphaD$  were introduced in \cite{Gleyzes:2015pma,Gleyzes:2015rua} to characterize conformal and disformal transformations that depend only on the scalar field value. 
Indeed, the structure of the action restricted to $\alphaH=\aL=\beta_A=0$ is invariant under this subset of transformations 
\cite{Bettoni:2013diz,Gleyzes:2015pma}. 
The function $\alphaX$ was introduced in  \cite{DAmico:2016ntq} to describe the transformation of the  action restricted only to $\aL=\beta_A=0$. The relations between  the effective parameters $\alphaK$, $\alphaB$, $\alphaT$, $\alphaM$ and $\alphaH$ in different frames were given in these references.

Here we extend these results to the general action \eqref{SBAction0}. 
In particular, as explicitly shown in App.~\ref{app:disformal}, the effective parameters in the quadratic action derived from $\tilde{S}$ are related to those associated with $S$ via the transformations:
\be \label{alphatilde}
\begin{split}
\tilde{M}^2 &=\frac{M^2 }{ C\sqrt{1+\alphaD}}\; , \\
\tilde{\alpha}_{\rm L}&=\aL \; , \\
\tilde{\alpha}_{\rm T}&=(1+\alphaT)(1+\alphaD)-1\; , \\
\tilde{\alpha}_{\rm H}&= \Xi\,  (1+\alphaD)\big[1+\alphaH-\alphaY (1+\alphaT)\big]-1\;,\\
\tilde{\bun}&=\Xi \big[ \alphaY(1+\aL)+\bun\big]\; ,\\
\tilde{\bdeux}&=\Xi^2\big[\bdeux-6\alphaY(\alphaY (1+\aL)+2\bun)\big]\; , \\
\tilde{\btrois}&=\Xi^2 (1+\alphaD) \big[\btrois+2 \alphaY^2(1+\alphaT)-4\alphaY(1+\alphaH)\big] \; ,
\end{split}
\ee
where we have introduced
\be
  \Xi \equiv  \frac{1}{(1+\alphaD)(1+\alphaX+\alphaY)} \,.
\ee
This function is always finite for an invertible transformation \eqref{disf_cov}. Indeed, in terms of the parameters \eqref{defalphas2},  the condition \eqref{invertible} implies $1+\alphaX +\alphaY \neq 0$, while  one must impose $1+\alphaD>0$ in order to conserve the metric signature.
Here we focus our attention on the parameters that are directly involved in the degeneracy constraints and do not show the analogous transformations for the other parameters $\alpha_{\rm K}$ and $\alpha_{\rm B}$, whose explicit expressions are given in App.~\ref{app:disformal}.
One can check that the two sets of degeneracy conditions (\ref{Ia}) and (\ref{IIa}), as well as the common condition (\ref{C_U}), are all invariant under the above disformal transformations. 

Interestingly, for degenerate theories satisfying either (\ref{Ia}) or (\ref{IIa}), it is possible to cancel simultaneously all three $\beta_A$ via a conformal transformation verifying
\beq
\alphaY=-\frac{\bun}{1+\aL}\,,
\eeq
%DL
provided $\aL\neq -1$.
One can also cancel $\aH$ via a disformal transformation such that 
\beq
\alphaX=\aH- (2+\aT)\alphaY\,.
\eeq
As a consequence, it is possible to cancel both $\beta_A$ and $\aH$ via a disformal transformation characterized by
\beq
\alphaY=-\frac{\bun}{1+\aL}\,, \qquad \alphaX=\aH+\frac{2+\aT}{1+\aL}\bun \;.
\eeq
Such a transformation is well defined for theories satisfying the conditions (\ref{Ia}) but not theories verifying (\ref{IIa}) for which the quantity $1+\alphaX+\alphaY$ vanishes. 
 In the first case, one simply recovers the property that 
 %DL
 %all the
 theories belonging to the same subclass as Horndeski can be related to Horndeski via a disformal transformation.

\section{Dispersion relation for a Minkowski background}
\label{sec:3}
For simplicity, we first study the linear perturbations about a  Minkowski background for theories of the form (\ref{action}). We thus specialize the quadratic action (\ref{SBAction0}) to the limit $a=1$ and $H=0$: this is equivalent to assuming that the typical frequencies and wave numbers  are much higher than the cosmological ones. For convenience, we redefine the coefficients of the $\d N^2$ and $\d K \d N$ terms as 
\beq
\label{MKB}
\MsqK=H^2\alphaK\,, \qquad \MB=H \alphaB\,,
\eeq
and we assume that the mass parameters $M_{\rm K}$ and $\MB$ can take any finite value in the Minkowski limit.

The scalar type perturbations in the unitary gauge can be expressed in terms of the quantities $\psi$ and $\zeta$, defined by
\beq
\label{UG}
 N^i=\d^{ij}\partial_j\psi\,,\qquad h_{ij}=e^{2\zeta}\d_{ij}\,. 
\eeq
Substituting  into the quadratic action, we thus obtain an action that depends on the three perturbations $\zeta$, $\delta N$, $\psi$ and their derivatives.  All the coefficients are constant since we are now in a Minkowski background.

\subsection{Dispersion relation and degeneracy}
In order to derive the dispersion relation, one considers perturbations of the form 
\beq
\begin{pmatrix}
\d N(t, {\bf x})\\
\zeta(t, {\bf x})\\
\psi(t, {\bf x})
\end{pmatrix}
=e^{-i\omega t+i{\bf k}\cdot{\bf x}}
\begin{pmatrix}
\d N(\omega, {\bf k})\\
\zeta(\omega, {\bf k})\\
\psi(\omega, {\bf k})
\end{pmatrix}\equiv e^{-i\omega t+i{\bf k}\cdot{\bf x}}\,  U\,,
\eeq
where $U$ denotes the column vector of the three perturbations in Fourier space. 
The resulting quadratic Lagrangian is of  the form 
\beq
L^{(2)}=U^\dagger K_U U\,,
\eeq
where $K_U$ is the $3\times 3$ kinetic matrix  with components
\be
\label{KM}
 K_U=M^2\begin{pmatrix}
\MsqK +   \bdeux \omega^2 + \btrois k^2   &  2(1+\aH)k^2+6\bun \omega^2+6 i \MB \omega  & 2(\MB  - i \bun \omega) k^2\\
 2(1+\aH)k^2+6\bun \omega^2-6 i \MB \omega &  -6(1+\aL) \omega^2 +2(1+\aT)k^2 &    2 i (1+\aL) \omega k^2\\
 2(\MB  + i \bun \omega) k^2  &- 2 i (1+\aL) \omega  k^2&- \frac23 \aL k^4
\end{pmatrix}  \;.
\ee
One finds the dispersion relation by imposing 
\beq
{\rm det} K_U=0\,,
\eeq
which yields 
\be
\begin{split}
\omskz\,  \omega^4  +\big(\omfkt k^2+ \omfkz \big)\omega^2 + \omtkf k^4+ \omtkt k^2 =0\; ,
\end{split}
\ee
with the  coefficients
\be
\begin{split}
\omskz = & \ 3 \big[(1+\aL)  \bdeux+ 6 \bun^2\big]\; ,\\
\omfkt = & \  6  \big[ 2 (1+\alphaH)+  (1+\alphaT)\bun\big]\bun +\aL  (1+\aT) \bdeux+3 (1+\aL) \btrois\;, \\
\omfkz = &\    3 \big[  (1+\aL)\MsqK+6\MB^2 \big] \; ,\\
\omtkf = & - \aL \big[2 (1+\alphaH)^2 - (1+\alphaT) \btrois \big]\; ,\\
\omtkt = & \  (1+\alphaT) \left( \aL \MsqK+6 \MB^2 \right)\;.\\
\end{split}
\ee
In the general case, the dispersion relation is a quartic polynomial in $\omega$ with only even powers, 
which means that there are two solutions for $\omega^2$, corresponding to the presence of two scalar modes, as expected. 
%% \beta_2: kineticity, \beta_3: braiding 
In particular, we note that the two parameters $\bun$ and $\bdeux$ contribute to the highest order coefficient in $\omega$, which is consistent with their interpretation given in the introduction. Interestingly, 
the structure of the coefficient $\omskz$ is the same as that of $\omfkz$ with $\bun$ and $\bdeux$  playing the role of $\alphaB$ and $\alphaK$, respectively (reminding that $M_{\rm B} \equiv H \alphaB$ and $M_{\rm K}^2 \equiv H^2 \alphaK$).
%the structure of the coefficients $\omskz$ and $\omfkz$ is the same with $\bun$ and $\bdeux$ respectively playing the role of $\alphaB$ and $\alphaK$ (reminding that $M_{\rm B} \equiv H \alphaB$ and $M_{\rm K}^2 \equiv H^2 \alphaK$). 
%playing the role respectively of $\MB$ and $\MsqK$. 
Note also that the highest term in spatial derivatives disappears when $\aL=0$.

If the condition $\omskz=0$ is satisfied, which is equivalent to the condition $\CU$ identified in (\ref{C_U}), then only a single scalar mode remains.
However, the above statement is valid in the unitary gauge for linear perturbations. In an arbitrary gauge, unless the theory is fully degenerate, i.e.~it satisfies the other conditions in (\ref{Ia}) 
%DL
or
 (\ref{IIa}), there still exists an extra scalar mode that simply does not show up at the linear level (see discussion in App.~\ref{app_degeneracy_2}). In order to ensure the absence of this extra scalar mode, 
it is sufficient to require either of the degeneracy conditions \eqref{Ia} and \eqref{IIa}, 
%% \aL\neq 0, k^4 -> breaking of Lorentz 
which implies $\omfkt=0$ and $\omtkf =0$. 
In this case, 
the dispersion relation takes the  very simple form 
\be
\label{disprelMin}
\omega^2- c_s^2 k^2=0, \qquad c_s^2\equiv - \frac13 \frac {(1+\alphaT)(6 \MB^2 + \aL \MsqK)}{\MsqK (1+\aL)+6\MB^2}\qquad ({\rm degenerate})\,.
\ee

\subsection{Newtonian limit}
One can also use the quadratic action to explore how the usual Poisson equation is modified. To do so, we consider the static limit (i.e. $\omega=0$) of the quadratic action and we introduce a point mass $m$ which is minimally coupled to the metric. 
If we work in the unitary gauge, we find that the kinetic matrix \eqref{KM} for the variables $\d N$, $\zeta$ and $\psi$  reads, in the limit $\omega=0$, 
\be
 K_V=M^2\begin{pmatrix}
   \btrois k^2 + \MsqK  &  2(1+\aH)k^2  & 2\MB k^2\\
 2(1+\aH)k^2 &  2(1+\aT)k^2 &    0\\
 2\MB k^2 & 0 &- \frac23 \aL k^4
\end{pmatrix}  \;.
\ee
The second line implies
\beq
\zeta=-\frac{1+ \alphaH}{1+\alphaT}\d N\,,
\eeq
which is equivalent to the relation
\beq
\Psi=\frac{1+ \alphaH}{1+\alphaT}\Phi\;,
\eeq
between the gravitational potential $\Phi$ (which in general
is related to the unitary gauge variables by $\Phi=\d N+\dot\psi$) and the spatial gravitational potential $\Psi=-\zeta$. 

We now distinguish the two cases $\aL\neq 0$ and $\aL=0$. 
If  $\aL\neq 0$, the last line of the kinetic matrix yields 
\beq
k^2\psi= 3\frac{\MB}{\aL}\d N\,.
\eeq
Substituting into the first line  and going back into real space, one finds the generalized Poisson equation
\beq
\label{Poisson}
M^2\left[2\frac{(1+\a_H)^2}{1+\a_T}-\btrois\right]\Delta\Phi + M^2\left(\MsqK+6\frac{\MB^2}{\aL}\right)\Phi =m\,  \d^{(3)}({\bf x})\,,
\eeq
where $\Delta\equiv \delta^{ij}\partial_i\partial_j$ denotes the Laplacian.
The coefficient in front of $\Delta\Phi$ in the generalized Poisson equation (\ref{Poisson}) corresponds to $(4 \pi G_{\rm N})^{-1}$, where $G_{\rm N}$ is the effective Newton constant. 
For  DHOST theories with $\aL\neq 0$, we see immediately that the coefficient  in front of the Laplacian in the Poisson equation vanishes, because of (\ref{IIa}), which means that  the effective Newton constant in the linear regime is infinite for these theories.

If  $\aL= 0$, one obtains the 
generalized Poisson equation
\beq
M^2\left[2\frac{(1+\a_H)^2}{1+\a_T}-\btrois\right]\Delta \Phi +M^2 \MsqK \Phi =m\,  \d^{(3)}({\bf x})\,.
\eeq
For DHOST theories that satisfy the conditions (\ref{Ia}) but not (\ref{IIa}), 
one thus gets a finite Newton constant $G_{\rm N}$
in the linear regime
\beq
\label{GNewton}
8 \pi G_{\rm N}= \frac{1}{M^2} \left[\frac{(1+\a_H)^2}{1+\a_T}-\frac{\btrois}{2} \right]^{-1}\,.
\eeq
These results seem to indicate that only theories  that are related to Horndeski via conformal or disformal transformations are phenomenologically viable. One should however 
%DL
investigate whether  
this peculiar behaviour persists at the nonlinear level.

We can check that for $\btrois=0$ this effective Newton constant agrees with the one found for Horndeksi and beyond Horndeski theories  in the quasi-static regime. In our notation, the full expression is given by eq.~(3.24) of \cite{DAmico:2016ntq} (see also \cite{Kobayashi:2014ida}).  In the absence of background matter and for $a$, $M$ and $\alphaH=$ constants, this reads
\be
8 \pi G_{\rm N}  = \frac{1}{2 M^2 (1+\alphaH)^2} \left[1 +\alphaT + \frac{2 \xi^2}{c_s^2 (\alphaK +6 \alphaB^2)} \right] \,,
\ee
with $\xi \equiv \alphaB (1+\alphaT) +\alphaT - \alphaM - \alphaH (1+\alphaM) $.
It is easy to check that, in the Minkowski limit, the last term in the brackets becomes  $1+\alphaT$ and  eq.~\eqref{GNewton} with $\btrois=0$ is recovered.
%

%%%%%%%%%%%%%%%%%%%%%%%%%%%%%%%%%%%%%%%%%%%%%
\section{Quadratic action in a cosmological background}
%%%%%%%%%%%%%%%%%%%%%%%%%%%%%%%%%%%%%%%%%%%%%
\label{sec:cosmo}

In this section,  we study the quadratic action for the  propagating degrees of freedom in a cosmological background.
 Since none of the additional operators studied here  contributes to tensor modes, we restrict our analysis to scalar perturbations. A derivation of the quadratic action for tensor perturbations, 
 which depends only on the parameters $M^2$ and $\aT$, 
  can be found for instance in \cite{Gleyzes:2013ooa,Gleyzes:2014rba}.
Using the usual expressions for the scalar perturbations in unitary gauge, 
\beq
\label{UG_FL}
 N^i=\d^{ij}\partial_j\psi\,,\qquad h_{ij}=a^2(t)\, e^{2\zeta}\d_{ij}\,,
\eeq
we can express the action~\eqref{SBAction0} as a functional of $\zeta$, $\delta N$ and $\psi$ and their derivatives. 
It is convenient to distinguish the two cases $\aL =0$ and $\aL \neq 0$.

\subsection{Case $\aL =0 $}

Let us assume $\aL=0$, i.e.~the first condition of eq.~\eqref{Ia}. Using eq.~\eqref{UG_FL} and noticing that the terms quadratic in $\Delta \psi$ cancel up to a total derivative, action \eqref{SBAction0} becomes
\be  
\begin{split}
\label{actionaLz}
S^\quadac = & \ \int d^3 x\,  dt \, a^3 \frac{M^2}{2}\bigg\{ -6 \dot \zeta^2 +12 \bun \dot \zeta \delta \dot N + \bdeux \delta \dot N^2 +12 H \left[ (1+\alphaB) \dot \zeta - \bun \delta \dot N \right] \delta N \\
& + H^2 (\alphaK - 6 - 12 \alphaB) \delta N^2 + 4 \left[ \dot \zeta - \bun \delta \dot N - H(1+\alphaB) \delta N \right] \Delta \psi \\
& + \frac{1}{a^2} \left[ 2 (1+\alphaT) (\partial_i \zeta)^2 + 4 (1+\alphaH) \partial_i \zeta \partial_i \delta N + \btrois \left(\partial_i \delta N\right)^2 \right] \bigg\} \;.
\end{split}
\ee
The kinetic Lagrangian is given by the first three terms on the right-hand side,
\be
\label{kinM1}
{\cal L}^\quadac_{\rm kin}  = a^3 \frac{M^2}{ 2 } (-6 \dot \zeta^2 +12 \bun \dot \zeta \delta \dot N + \bdeux \delta \dot N^2) \;.
\ee
Thus, without any assumption on the time dependent functions $\beta_A$, the action above  describes in general two propagating scalar modes, $\zeta$ and $\delta N$, while $\psi$, which appears without time derivatives, can be treated as a Lagrange multiplier. 
In this general case, the full analysis, including matter, extends our analysis of the previous section and is discussed in App.~\ref{app:matter}.

The expression (\ref{kinM1}) is degenerate if the determinant of the kinetic matrix vanishes, i.e.~if 
\be
\bdeux=-6 \bun^2 \;,
\ee
where one recognizes the second condition in eq.~\eqref{Ia}, or the condition \eqref{C_U} with $\aL=0$.
 In this case the kinetic Lagrangian can be written as 
\be
{\cal L}^\quadac_{\rm kin}  = -6 a^3 \frac{M^2}{ 2 }  ( \dot \zeta - \bun  \delta \dot N )^2 \;,
\ee
which suggests that the time derivatives of $\delta \dot N$ can be eliminated by replacing the variable $\zeta$ 
%DL
%by 
with
the new variable
\be
\tzeta \equiv \zeta - \bun \delta N \;.
\ee
This variable describes the propagating scalar degree of freedom in the degenerate case.

To find the associated  quadratic action, we can proceed similarly to what was done in \cite{Gleyzes:2013ooa,Piazza:2013coa}.
Varying the action with respect to $\psi$ yields the scalar component of the momentum constraint. 
In terms of the new variable, it reads
\be
\label{momaz}
\delta N=\frac{\dot \tzeta}{H(1+\alphaB)-\dot \bun} \; .
\ee

Substituting this expression for  $\delta N$ into the action
  and performing some integration by parts,
   we finally find the quadratic action for the propagating degree of freedom $\tzeta$, 
\be
\label{actzIa}
S^\quadac= \int d^3 x \, dt \, a^3   \frac{M^2 }{2} \bigg[  {\Az}_{\tzeta}   \dot{\tzeta}^2 +  {\Bz}_{\tzeta} \frac{(\partial_i \tzeta)^2}{a^2} +  {\Cz}_{\tzeta} \frac{(\partial_i \dot \tzeta)^2}{a^2}  \bigg] \; .
\ee
Here ${\Az}_{\tzeta}$, ${\Bz}_{\tzeta}$ and ${\Cz}_{\tzeta}$ are background-dependent 
functions  whose explicit expressions are
\begin{align}
{\Az}_{\tzeta} & = \frac{  1}{(1+\alphaB-\dot \bun /H)^2}  \bigg[\alphaK+ 6\alphaB^2 - \frac{6}{a^3 H^2 M^2} \frac{d}{dt} \left( a^3 H M^2 \alphaB \bun \right) \bigg]   \; , \label{AzI}\\
{\Bz}_{\tzeta} & =  2 (1+\alphaT)-\frac{2}{a M^2 }\frac{d}{dt}\bigg[\frac{a M^2 \big( 1+\alphaH+\bun(1+\alphaT)\big)}{H(1+\alphaB)-\dot \bun}\bigg]   \; , \label{BzI} \\
{\Cz}_{\tzeta} & = \frac{4(1+\alphaH) \bun + 2 (1+\alphaT) \bun^2 + \btrois}{(1+\alphaB-\dot \bun /H)^2} \;.
\end{align}

As the above action implies that $\tzeta$ is conserved in the long wavelength limit, i.e.~$\dot \tzeta \approx 0$ for $k\ll aH$,  and eq.~\eqref{momaz} implies that $\delta N$ vanishes in the same limit, it follows that $\zeta$  is conserved on large scales, 
\be
\dot \zeta \approx 0 \qquad  (k \ll aH)  \;,
\ee
as in the more standard case of Horndeski and beyond Horndeski theories \cite{Gleyzes:2014rba}.

Up to now we have imposed only the first two  conditions in  eq.~\eqref{Ia}. 
  Since only $\bun$ enters in the above expressions for ${\Az}_{\tzeta}$ and ${\Bz}_{\tzeta}$, these definitions remain unchanged when the full degeneracy conditions $\CI$ are imposed, while the function $\Cz_{\tzeta}$ vanishes.
In this case the action takes the usual form,
\be
\label{actionzetaCI}
S^{\quadac}= \int d^3 x\,  dt\,  a^3   \frac{M^2 }{2} \bigg[  {\Az}_{\tzeta}   \dot{\tzeta}^2 +  {\Bz}_{\tzeta} \frac{(\partial_i \tzeta)^2}{a^2}  \bigg] \;.
\ee
Absence of instabilities requires that 
%DL
the  coefficients in the action satisfy
\be
{\Az}_{\tzeta} \ge 0 \;, \qquad {\Bz}_{\tzeta} \le 0 \; .
\ee
In particular, the first condition is equivalent to 
\be
\label{alphaZ}
\alpha\equiv \alphaK+ 6\alphaB^2 - \frac{6}{a^3 H^2 M^2} \frac{d}{dt} \left( a^3 H M^2 \alphaB \bun \right)\geq 0 \,.
\ee

The dispersion relation is standard, $\omega^2 = c_s^2 k^2/a^2$, with a 
sound speed  given by
\be
\label{cs0Ia}
c_s^2 = -{{\Bz}_{\tzeta}}/{{\Az}_{\tzeta}} \;.
\ee
It is straightforward to check that for $\bun=0$ one recovers the  sound speed derived for theories belonging to the Horndeski (see eg.~\cite{DeFelice:2011bh}) and beyond Horndeski classes in \cite{Gleyzes:2013ooa,Gleyzes:2014qga,Gleyzes:2014rba}. 
Moreover, in the Minkowski limit, we recover eq.~(\ref{disprelMin}) in the case $\aL=0$. 
One can also verify that under a general transformation (\ref{disf_cov}),  the above expression for the sound speed  transforms like the lightcone, i.e.~$ \tilde c_s^2 =  (1+\alphaD)c_s^2 $, as expected.

\subsection{Case $\aL \neq 0$}
We now consider the case $\aL \neq 0$. Varying 
the action with respect to $\psi$ yields the scalar component of the momentum constraint, which reads
\be
\label{costrpsi}
\frac{\aL}{3} \,\frac{\Delta \psi}{a^2}  +    (1+ \aB+  \aL)  H \delta N+ \bun {\delta \dot N}-  (1+\aL) \dot{\zeta}=0\; .
\ee
Using this expression to eliminate $\psi$ from the action,
  one obtains
\be\label{actionConstrPsi}
\begin{split}
S^\quadac =& \ \int d^3 x\,  dt \, a^3 \frac{M^2}{ 2 }  \bigg\{ \frac{1}{  \aL }\Big[ 6 (1+\aL)  \dot \zeta ^2 -1 2 \bun   \delta \dot N  \dot \zeta + (6 \bun^2  + \aL \bdeux )  \delta \dot N ^2   \\
& +12   H \left( \bun  \left(1 + \aB \right)  \delta \dot N - \left(1+ \aB +  \aL \right) \dot \zeta \right)  \delta  N  \\
& +H^2 \Big(6  \aB^2+12  \aB + \aL  \alphaK  +6  (1+\aL) \Big)  \delta  N ^2\Big] \\
&  +\frac{1}{ a^2}  \left[ 2 \left(1+  \alphaT \right) \left(\partial_i  \zeta \right)^2 +4 \left(1+ \alphaH\right) \partial_i \zeta \, \partial^i \delta  N   + \btrois  \left(\partial_i  \delta N\right)^2 \right] \bigg\} \;.
\end{split}
\ee
Once again, in the absence of any assumption on the $\beta_A$,  this action describes in general two propagating scalar modes, $\zeta$ and $\delta N$. 

The kinetic matrix is degenerate for 
\be
\label{uni_degeneracy}
 \bdeux=-6\frac{\bun^2}{1+\aL} \;,
\ee
which corresponds to the condition of eq.~\eqref{C_U}. 
It can be diagonalized by introducing  the variable 
\be\label{tzetadef}
\tzeta=\zeta-\frac{\bun}{1+\aL} \delta N \; ,
\ee
which represents the propagating degree of freedom in this case.

Using these two relations in action \eqref{actionConstrPsi} and performing some integrations by parts, the action takes the form
\be \label{Sxyx}
\begin{split}
S^\quadac =& \ \int d^3x \, dt\, a^3 \frac{M^2}{2}\,  \bigg\{   6 \frac{ 1+\aL}{\aL} \dot \tzeta ^2 + 2 \,\VVV_{12} \,\dot \tzeta \delta N +    \MMM_{22} \, \delta N ^2 \\ 
& + \frac{1}{a^2}  \left[2(1+\alphaT) (\partial_i \tzeta)^2  +2 \SSS_{12}\,  \partial_i \tzeta\,  \partial^i \delta N+ \SSS_{22}  (\partial_i \delta N)^2 \right]  \bigg\} \; , 
\end{split}
\ee
with
\be
\begin{split}
\SSS_{12}  &  =  2 \Big[1+\alphaH+\bun\frac{1+\alphaT}{1+\aL}\Big]\;, \\
\SSS_{22} & = 4\bun \frac{1+\alphaH}{1+\aL}+2\bun^2\frac{1+\alphaT}{(1+\aL)^2}+\btrois \;.
\end{split}
\ee
The explicit expressions of $\VVV_{12}$ and $\MMM_{22}$ above are not relevant for this discussion and can be found in App.~\ref{app:elements}.
Variation of action~\eqref{Sxyx} with respect to $\delta N$ gives then the Hamiltonian constraint equation, 
\be\label{constrDelNsol}
\VVV_{12} \dot \tzeta + \MMM_{22} \delta N  =  \frac{1 }{a^2} \left[ \SSS_{12} \Delta \tzeta + \SSS_{22}  \Delta \delta N \right] \; .
\ee
The constraint equation \eqref{constrDelNsol} can be solved for $\delta N$ as a function of $\tzeta$ and $\dot \tzeta$ and plugged back into eq.~\eqref{Sxyx}, which 
yields the quadratic action for the propagating degree of freedom $\tzeta$. 
If $\SSS_{12}$ and $\SSS_{22}$ 
do not vanish, this action contains higher spatial derivatives and  Laplacian operators in the denominator, resulting in a nonstandard dispersion relation for $\tzeta$.
We have written its explicit expression in App.~\ref{app:matter}, including also matter perturbations for completeness. Analogously to what happens in the previous section, there is no mass term in the action and thus $\tzeta$ and $\zeta$ are conserved in the long wavelength limit.

We now
derive the action for $\tzeta$ when the full set of degeneracy conditions \eqref{IIa} is imposed. In this case, $\SSS_{12} = 0 = \SSS_{22}$ and the constraint equation \eqref{constrDelNsol} simplifies,
\be
\delta N = -\frac{\VVV_{12}}{\MMM_{22}} \dot \tzeta \;  \qquad (\CII) \;,
\ee
while the expressions for $\VVV_{12}$ and  $\MMM_{22}$ take the following form 
\be \label{VMdef}
\begin{split}
\VVV_{12}  =&\ -\frac{6} {  \aL}\bigg[H  (1+ \alphaB+\aL)+(1+\aL) \frac{d}{dt} \left(\frac{1+\alphaH}{1+\alphaT}\right) \bigg]  \;  \qquad (\CII)   \; , \\
\MMM_{22}=& \  \frac{\aL }{6 (1+\aL)} \VVV_{12}^2+ H^2 \left( \alphaK +  \frac{6 \alphaB^2}{1+\aL} \right)+ \frac{6}{a^3 M^2 } \frac{d}{dt}\left(a^3 M^2 H \alphaB \frac{1+\alphaH}{1+\alphaT} \right) \;  \qquad (\CII) \; .
\end{split}
\ee
Replacing $\delta N$ into  action \eqref{Sxyx} using the above constraint,  one obtains a quadratic action of the form of eq.~\eqref{actionzetaCI}, with
\be
A_{\tzeta}  =  6 \frac{ 1+\aL}{\aL}-\frac{\VVV_{12}^2 }{\MMM_{22}} \;, \qquad B_{\tzeta} = 2 (1+\alphaT)\; \qquad (\CII) \;.
\ee

In the limit $\aL = 0$ these expressions  reduce to those in eqs.~\eqref{AzI} and \eqref{BzI}, with the function $\bun$ given by eq.~\eqref{IIa}, i.e.
\be
A_{\tzeta}  =  \frac{\alphaK + 6 \alphaB^2 + \frac{6}{a^3 H^2 M^2} \frac{d}{dt} \left( a^3 H M^2 \alphaB \frac{1+\alphaH}{1+\alphaT} \right) }{ \left[1+\alphaB + \frac{d}{dt}\left( \frac{1+\alphaH}{1+\alphaT} \right) \right]^2}  \;, \qquad B_{\tzeta} = 2 (1+\alphaT)\; \qquad (\CI \cap \CII) \;.
\ee

%%%%%%%%%%%%%%%%%%%%%%%%%%%%%%
\section{Lorentz-breaking theories}
\label{sec:5}
%%%%%%%%%%%%%%%%%%%%%%%%%%%%%%
In this section, we show how our analysis can also be applied to  theories already proposed in the literature, 
such as Lorentz-breaking theories 
%DL
inspired by 
Horava's gravity and khronometric theories.

\subsection{Lorentz-breaking theories}
In our analysis of DHOST theories, we have started from a covariant formulation of the action and then derived the Lagrangian in the unitary gauge. 
Several  models that explicitly break Lorentz invariance have been proposed in the literature and their action is  often given directly in the unitary gauge. An illustrative example is 
Horava gravity\cite{Horava:2009uw}, with several of its extensions.  A general presentation of these models can be found in \cite{Blas:2010hb}, which we will follow in our discussion below.  The actions studied in  \cite{Blas:2010hb} are of the form\footnote{The notation for  the coefficients in the action differs from the one of \cite{Blas:2010hb}.}
\beq
\begin{split}
\label{S_u}
S_{\rm unitary}=&\int d^3x\,  dt \sqrt{\h} N\Bigg\{\frac{M^2}{2}\left(K_{ij}K^{ij}-\lambda K^2
+\frac{\lambda_2}{N^6} \left(\dot N- N^i\partial_i N\right)^2-\lambda_3
\frac{\partial_i N \partial^iN}{N^2}-{\cal V}\right)
%\right.
\\
&
%\left.
\qquad\qquad\qquad
+ \frac{\lambda_0}{4}\mu^4\left(\frac{1}{N^2}-1\right)^2\Bigg\}\,,
\end{split}
\eeq
and ${\cal V}$ is some potential term which is not relevant for us and will be ignored below. As one can see, the action contains an explicit dependence on $\dot N$. 
In \cite{Blas:2010hb}, different cases are considered, depending on the values of the parameters in the Lagrangian. The case $\lambda_2=0$ and $\lambda_3=0$ corresponds to  Horava's projectable  model in the limit $\lambda_0\rightarrow \infty$. The healthy extension of Horava's non projectable model \cite{Blas:2009qj} corresponds to $\lambda_0=\lambda_2=0$. Finally,   \cite{Blas:2010hb} also considers an extension with $\lambda_0$  and $\lambda_2$ non zero. 

It is always possible to rewrite the unitary action (\ref{S_u}) as a covariant action by using the Stueckelberg formalism, where the time coordinate of the unitary gauge is replaced by a scalar field $\phi$. This leads to the covariant action~\cite{Blas:2010hb}
\bea
S_{\rm cov}&=&\int d^4x \sqrt{- g}\Bigg\{ 
\frac{M^2}{2}\bigg[ {}^{(4)}\!R
+\frac{\lambda-1}{X}\left(\square\phi-\frac{1}{X}\phi^\mu\phi_{\mu\nu}\phi^\nu\right)^2+\frac{\lambda_2}{\mu^4X^2}\left(\phi^\mu\phi_{\mu\nu} \phi^\nu\right)^2
%\right.
%\right.
\cr
&&\qquad\qquad \quad 
%\left.
%\left.
+\frac{\lambda_3}{X^2}\left(\phi^\mu\phi_{\mu\nu}\phi^{\nu\lambda}\phi_\lambda-\frac{1}{X} \left(\phi^\mu\phi_{\mu\nu}\phi^\nu\right)^2\right)
\bigg]+\frac{\lambda_0}{4\mu^4}(X+\mu^4)^2
\Bigg\}\,.
\label{BPS_cov}
\eea
One immediately sees that this action is of the form (\ref{action}) with no cubic terms and
\beq
\begin{split}
\aq_1=&\, 0\,, \qquad \aq_2=\frac{(\lambda-1)M^2}{2X}\,,\qquad \aq_3 = \frac{(1-\lambda)M^2}{X^2}\,, \\
\aq_4=&\, \frac{\lambda_3M^2}{2X^2} \,, \qquad \aq_5=M^2\left(\frac{\lambda-\lambda_3-1}{2X^3}+\frac{\lambda_{2}}{2\mu^2 X^2}\right)\,,
\end{split}
\eeq
$P(X)=\lambda_0 (X+\mu^4)^2/(4\mu^4)$ and $Q(X)=0$.
This corresponds to the following effective parameters
\beq
\aT=\aH=0, \qquad \aL=3(\lambda-1)/2\,, \quad \bun=0\,, \quad \bdeux= \lambda_2\,, \quad \btrois= \lambda_3\,.
\eeq
and $\aK=\lambda_0\mu^4/(2H^2 M^2)$.
%, \aB ???

When  $\lambda_2=0$, one has $\bun=\bdeux=0$ and the condition (\ref{C_U}) is verified. This means that one finds a single propagating scalar mode in the unitary gauge. However, even if $\lambda_2=0$,  the parameters do not satisfy the full degeneracy 
conditions, unless $\lambda_3=2$. 
The covariant theory (\ref{BPS_cov}) (with $\lambda_2=0$) is not a DHOST theory and thus contains an extra degree of freedom, 
although it is not directly visible in the unitary gauge
%DL
(see discussion in App.~\ref{app_degeneracy_2}).
This is an example of theory that looks degenerate in the unitary gauge but is not degenerate. 
In order to get rid of this dangerous extra degree of freedom, one can either define the theory directly in the unitary gauge or  consider the covariant action but only for a restricted range of solutions, as discussed in \cite{Blas:2010hb}.

\subsection{Khronometric theories}

In khronometric theories \cite{Blas:2010hb}, the scalar field Lagrangian is invariant under the field redefinition 
$\phi \to \tilde \phi (\phi )$.
At lowest order in derivatives, the action can be written in terms of the unit vector field 
 $u_\mu \equiv {\partial_\mu \phi}/{\sqrt{-X}}$ as
\be
\label{ackhron}
S= \frac{\Mp^2}{2} \int d^4 x \sqrt{-g}  \left( {}^{(4)}\!R+ {\cal K}^{\mu \nu}_{\ \ \rho \sigma} \nabla_\mu u^\rho \nabla_\nu u^\sigma \right) \;,
\ee
with 
\be
{\cal K}^{\mu \nu}_{\ \ \rho \sigma} \equiv c_1 g^{\mu \nu}g_{\rho \sigma} + c_2 \delta_\rho^\mu \delta_\sigma^\nu + c_3 \delta^\mu_\sigma \delta_\rho^\nu +c_4 u^\mu u^\nu g_{\rho \sigma} \;,
\ee
where $\Mp$ is a constant with mass dimension and the $c_i$ are dimensionless constants.
Without loss of generality, we can set $c_1=0$ in the action \cite{Creminelli:2012xb}.
Moreover, by using the Gauss-Codazzi relation and that $\nabla_\mu u_\nu = K_{\mu \nu} - u_\mu u^\rho \nabla_\rho u^\nu$, the above Lagrangian can be rewritten as
\be
S_{\rm unitary}= \frac{\Mp^2}{2} \int d^3 x\, dt \, \sqrt{h}\,  N \left[  R+ (c_2 -1) K^2 + (c_3+1) K_{ij} K^{ij}    + c_4 \frac{\partial_i N \partial^iN}{N^2}  \right] \;.
\ee
In the following we will assume that $c_3  > -1$, which ensures that 
\be
M^2 = \Mp^2 (1+c_3) > 0 \;, 
\ee
i.e.~that gravitons have  a strictly positive kinetic term.
Expanding at second order and comparing with eq.~\eqref{SBAction0}, one finds
\be
\alphaT = \alphaH = - \frac{c_3}{1+c_3} \;, \qquad \aL = - \frac32 \frac{c_2+c_3}{1+c_3} \;, \qquad \bun = \bdeux=0 \;, \qquad \btrois = \frac{c_4}{1+c_3} \;.
\ee

To compute the action for the propagating scalar degree of freedom, we can use action \eqref{actionConstrPsi} with $\alphaH=\alphaT$ and $\bun = \bdeux=0$. 
%Notice that there is no need to shift $\zeta$ by a term proportional to $\delta N$ as in eq.~\eqref{tzetadef}, because $\delta N$ appears without time derivatives.  Thus, variation with respect to $\delta N$ yields a constraint,
%\be
%6(1+\aL) H (\dot \zeta - H \delta N) + \frac1{a^2} \left[2 (1+\alphaT) \Delta \zeta+ \btrois \Delta \delta N \right]=0\;.
%\ee
 Using the Hamiltonian constraint to replace $\delta N$,
%this relation into action \eqref{Skhr1} 
and performing an integration by parts, the quadratic action for $\zeta$  can be written in Fourier space as
\be
\begin{split}
\label{ackhron2}
S^\quadac = & \ \int \frac{d^3 k}{(2\pi)^3}\,  dt \, a^3 \frac{ M^2}{2} \frac{k_H^2}{6(1+\aL) + \aL \btrois k_H^2 } \bigg\{ 6 (1+\aL) \btrois   \dot \zeta_{\bf k} \dot \zeta_{- \bf k} \\
& + \frac{2 a^2 H^2 (1+\alphaT)}{6(1+\aL) + \aL \btrois k_H^2} \left( c_{\zeta 2}  + c_{\zeta 4} k_H^2+ c_{\zeta 6}  k_H^4\right) \zeta_{\bf k} \zeta_{- \bf k} \bigg\} \;,
\end{split}
\ee
where 
\be
\begin{split}
c_{\zeta 2} &= 36 (1+\aL)^2 \dot H \;, \qquad c_{\zeta 4} = - 6 \aL (1+\aL)  \bigg[ \btrois \bigg(1+\frac{\dot H}{H^2} \bigg)+ 2 (1+\alphaT) \bigg] \;, \\
c_{\zeta 6} &= \aL^2 \, \btrois \left[\btrois - 2 (1+\alphaT)  \right] \;,
\end{split}
\ee
and $k_H \equiv k/(aH)$. This action can be also derived from \eqref{SxyxM} in App.~\ref{app:matter}.

If we now restrict our discussion to degenerate khronometric theories, the degeneracy conditions \eqref{Ia} and \eqref{IIa} impose constraints on the parameters of actions \eqref{ackhron2} and \eqref{ackhron}. We find, respectively, 
\be
\label{kh1}
\aL  =  0\,, \qquad \btrois= 0 \;, \qquad \Leftrightarrow \qquad c_2 + c_3  =  0\,, \qquad c_4= 0 \;, \qquad (\CI)\;, 
\ee
for conditions \eqref{Ia}, and 
\be
\label{kh2}
\aL  = -1 \,, \qquad \btrois= 2 (1+\alphaT) \;, \qquad \Leftrightarrow \qquad 3c_2  = 2 -c_3 \,, \qquad c_4= 2  \;, \qquad (\CII) \,,
\ee
for conditions \eqref{IIa}.
These two families of degenerate khronometric theories were already identified in \cite{Achour:2016rkg}, together with  two other families that lead to $M^2=0$ and are thus not relevant here.

As shown in \cite{Achour:2016rkg}, the first family (\ref{kh1}) is conformally-disformally related to general relativity, which means that there is no dynamical scalar degree of freedom in the absence of matter. 
The second family (\ref{kh2})  is conformally-disformally related to the theory (in units where $M_*^2/2=1$)
\be
f_2=1,\quad \aq_1=-\frac2X\,, \quad \aq_2=\aq_3=0, \quad \aq_4=\frac{6}{X^2}\,, \quad \aq_5=-\frac{4}{X^3}\,.
\ee
Substituting into the unitary gauge expression (\ref{LL2}), one finds
\be 
{\cal L} =  R  + 3 K_{ij} K^{ij} - K^2  +2 \frac{\partial_i N\partial^iN}{N^2}\;,
\ee
and the trace part of $K_{ij}$ automatically cancels, which means that there is no propagating scalar mode. Note that theories of this family possess the peculiar property that  $\bun=\bdeux=0$ while $\btrois\neq 0$. This is not in contradiction with our discussion in Sec. \ref{sec:2.3} as $\aL=-1$ here.

 \section{Conclusions}
In this work, we have studied the effective description of quadratic and cubic  Higher-Order Scalar-Tensor theories, focusing in particular on the degenerate ones, i.e. DHOST theories. 
We considered  the quadratic action of linear perturbations about a homogeneous and isotropic background, written in the unitary gauge where the scalar field is spatially uniform. In general, the quadratic action contains time and spatial derivatives of the lapse perturbation, which requires the introduction of three new (time-dependent) parameters which we denote $\bun$, $\bdeux$ and $\btrois$. 
We also need  another parameter, $\aL$, in front of the  trace of the extrinsic curvature (squared), in order to cover all DHOST theories.

 The presence of time  derivatives of the lapse is not in contradiction with the property that  DHOST theories contain a single scalar degree of freedom, because of the existence of degeneracy conditions that ensure that the effective parameters $\bun$, $\bdeux$ and $\btrois$ in the quadratic action are not arbitrary but instead must be linked via three relations. One of these relations, (\ref{C_U}), can easily be inferred from the requirement that the unitary gauge action is 
 manifestly
 degenerate, but the other two relations cannot be immediately guessed within the unitary gauge, because they come from the degeneracy imposed at the level of the covariant action. However, as we have shown, they can be deduced  from the requirement that the dispersion relation is linear.
 
 Remarkably, all cases can be summarized by only two different sets of degeneracy conditions at the level of the linear perturbations in the unitary gauge. The first set, (\ref{Ia}), is characterized by the condition $\aL=0$, while $\bun$ is left arbitrary and the other two  coefficients $\bdeux$ 
 and $\btrois$ are constrained in terms of $\bun$, $\aT$ and $\aH$. By contrast, with the alternative set of degenerate conditions (\ref{IIa}), the parameter $\aL$ remains arbitrary while the three coefficients $\bun$, $\bdeux$ and $\btrois$ are determined in terms of $\aL$, $\aT$ and $\aH$. 
 Among all quadratic and cubic DHOST theories, the subclass containing Horndeski, beyond-Horndeski and the  theories conformally-disformally  related to them stands out as satifying (\ref{Ia}) only. By contrast, all the other subclasses verify the conditions (\ref{IIa}), including 
 %DL
 one  that satisfies $\aL=0$ as well, i.e. both $\CI$ and $\CII$ (see Table \ref{table_DHOST}). 
 
 Let us turn to  phenomenology. 
  Our analysis shows that all DHOST subclasses, except the one containing Horndeski and beyond-Horndeski, suffer from the problem that the   effective Newton's constant becomes infinite in the linear regime. This analysis, which is restricted to linear perturbations, thus seems to indicate that one cannot recover standard gravity in these  theories, 
  although a fully nonlinear treatment would be necessary for a definite conclusion. 
Even if they cannot account for gravity as we know it, such theories could still be interesting for other contexts, such as in the early Universe.
  
  In cosmology, we derived  the quadratic action governing the dynamics of the linear scalar mode. 
  By imposing only the  condition $\CU$, defined in  (\ref{C_U}), we obtained an action with a nonlinear dispersion relation and we checked that the curvature perturbation $\zeta$ is conserved on large scales. When the full degeneracy conditions are imposed, the dispersion relation  simplifies and becomes linear. 
 
  We have also studied the impact of a fully general conformal-disformal transformation, with both functions depending on $X$ and $\phi$, on the quadratic 
  %DL
  perturbative
  action, thus extending previous results on the transformation of the effective parameters.  
 This expresses, at the level of linear perturbations, the underlying structure of transformations within DHOST theories, which enlarges those for Horndeski theories, where $C$ and $D$ are restricted 
 %DL
 to be independent of $X$, and those for beyond Horndeski, where $D$ can depend on $X$ but not $C$. 
  
  In the main body of this paper, we have implicitly assumed that matter is minimally coupled to the metric.
 In this case,  two conformally-disformally related theories are not physically equivalent. 
 %DL
 It is also possible to relax the assumption of minimal coupling
 % To establish a correspondence between two formulations that are physically equivalent,
%  Of course, one can also consider a conformal-disformal transformation acting on both the gravitational sector and the  matter sector. In this case, 
 % one must allow for a non-minimal coupling of matter to the metric. One can 
 and 
 attribute to each matter species four parameters that characterize, at the linear level, the non-minimal coupling to the metric. These four parameters transform under a general conformal-disformal transformation, similarly to the  
 %DL
 parameters of the scalar-tensor action. 
 In  App.~\ref{app:matter}, we derived the dispersion relation for the scalar mode in the presence of  matter, which enables to verify that the dispersion relation remains the same in all frames. 
Moreover,  the  mixing between matter and scalar perturbations can be quantified with a frame-invariant parameter. 
If we consider $N_S$ matter species, their coupling to the metric is described by $4N_S$ parameters, which add to the $6$ independent parameters for the gravitational sector (9 parameters minus three degeneracy constraints). Taking into account  frame transformations,  characterized by $4$ parameters, we end up with $2(2N_S+1)$ physically independent parameters.
  
 Note that one can use the disformal transformations  to simplify the scalar-tensor action, 
 %DL
 although at the price of complicating the coupling between matter and the metric.
  For example, for DHOST theories in the first subclass, one can use a conformal-disformal transformation to rewrite the scalar-tensor sector as a Horndeski theory, i.e.~with  $\aL=0$, $\aH=0$ and  $\bun=\bdeux=\btrois=0$, while matter has a complicated coupling to the metric and scalar field. This choice between the Jordan frame (where matter is minimally coupled) and the Horndeski frame is analogous to the choice between the Jordan frame and the Einstein frame for ordinary scalar-tensor theories. 

Finally, we have also discussed Lorentz-breaking theories inspired by Horava's gravity, as well as khronons. In particular, for the former, we have shown that the covariant 
formulations of these models lead to nondegenerate theories. 
Therefore, they contain an extra degree of freedom even if it is not visible at the linear level in the unitary gauge; see discussion in App.~\ref{app_degeneracy_2}.

\vspace{5mm}
\noindent
{\bf Acknowledgements}.
M.M.~and F.V.~acknowledge financial support from ``Programme National de Cosmologie and Galaxies'' (PNCG) of CNRS/INSU, France. We thank the  Galileo Galilei Institute, Florence, and the organizers of the  Workshop  ``Theoretical Cosmology in the Era of Large Surveys'' 2-6 May 2016, where this work was initiated. 

%%%%%%%%%%%%%%%%%%%%%%%%%%%%%%%%%%%%%%%%%%%%%%%%%%%%%%%%%%%%%%%%%%%%%%%
%%%%%%%%%%%%%%%%%%%%%%%%%%%%%%%%%%%%%%%%%%%%%%%%%%%%%%%%%%%%%%%%%%%%%%%%%
\appendix

\section{$(3+1)$ decomposition of scalar-tensor Lagrangians in the unitary gauge}
\label{app_unitary}

In the unitary gauge, specified by eq.~\eqref{UGphi}, each of the elementary Lagrangians defined in eqs.~\eqref{QuadraticL} and \eqref{CubicL} can be expressed in terms of  the 
velocities
$\Vs$ and $K_{ij}$. For the quadratic Lagrangians, eq.~\eqref{QuadraticL}, one finds
 \be
\begin{split}
L_1^\2=&\ \As  ^2 K_{ij} K^{ij} +\Vs^2-2 \partial_i\As   \partial^i \As   \;, \\
L_2^\2=&\ (\As  K+\Vs)^2\,,  \\
L_3^\2=& -\As  ^2 \Vs(\As  K+\Vs) \;, \\
L_4^\2=& \ \As  ^2\left(-\Vs^2+ \partial_i\As   \partial^i \As  \right)\,, \\ 
L_5^\2= & \ \As  ^4 \Vs^2 \;.
\end{split}
\ee
Moreover, the cubic Lagrangians in eq.~\eqref{CubicL} read
\be
\begin{split}
L_1^\3=& -(\As  K+\Vs)^3\;, \\
L_2^\3=&-(\As  K+\Vs) (\As  ^2 K_{ij} K^{ij} +\Vs^2-2 \partial_i\As   \partial^i \As   ) \;, \\
 L_3^\3=&-\As  ^3 K_j^i K_{il} K^{l j}+3 \As   K^{ij}  \partial_i\As   \partial_j \As  +3 \Vs  \partial_i\As   \partial^i \As  -\Vs^3\;,\\
L_4^\3=& \ \As  ^2 \Vs \,(\As  K+\Vs)^2 \;,\\
L_5^\3=&-\As  ^2\left(-\Vs^2+ \partial_i\As   \partial^i \As  \right)(\As  K+\Vs) \;,\\
L_6^\3=& \ \As  ^2 \Vs \,\left(\As  ^2 K_{ij} K^{ij} +\Vs^2-2 \partial_i\As   \partial^i \As  \right) \;,\\
L_7^\3=&-\As  ^3 K^{ij}  \partial_i\As   \partial_j \As  -2\As  ^2 \Vs \partial_i\As   \partial^i \As  +\As  ^2 \Vs^3 \;,\\
L_8^\3=& \ \As  ^4 \Vs\left(-\Vs^2+ \partial_i\As   \partial^i \As  \right)\;,\\
L_9^\3=&- \As  ^4 \Vs^2(\As  K+\Vs) \;,\\
L_{10}^\3=&\ \As  ^6 \Vs^3\;. \\
\end{split}
\ee

The terms that depend on the Ricci tensor can also be expressed in the unitary gauge. The simplest way to do this is to rewrite them 
in terms of the quartic and quintic Horndeski Lagrangians, respectively defined as
 \begin{align}
L_4^{\rm H} & =  f_2 {}^{(4)}\!R - 2 f_{2X} (\Box \phi^2 - \phi^{\mu\nu} \phi_{\mu\nu}) \,,
\label{quarticH}
\\
L_5^{\rm H} & =  f_3 {}^{(4)}\!G_{\mu\nu} \phi^{\mu\nu} + \frac{1}{3} f_{3X}(\Box \phi^3 - 3 \Box \phi \phi_{\mu\nu} \phi^{\mu\nu}
+ 2\phi_{\mu\nu} \phi^{\mu\sigma} \phi^\nu_{\, \sigma}) \,. \label{quinticH}
\end{align}
In our terminology, they correspond to a quadratic and a cubic Lagrangian, respectively.  Indeed, they are of the form \eqref{action}, with 
\be
a_1= - a_2 =  2 f_{2X} \,, \qquad a_3 = a_4 = a_5 =0 \,,
\ee
and 
\be
3 b_1= - b_2 =\frac32 b_3 = f_{3X} \,, \qquad  b_i = 0 \,\,\,(i=4,...,10) \,.
\ee
Therefore, the full action \eqref{action} can be rewritten as
\bea\label{action with Horndeski}
S[g,\phi] = \int d^4x \, \sqrt{ -g } \, \left( L_4^{\rm H}  +
 \tilde C_\2^{\mu\nu\rho\sigma} \,  \phi_{\mu\nu} \, \phi_{\rho\sigma}
 +  L_5^{\rm H} + 
\tilde C_\3^{\mu\nu\rho\sigma\alpha\beta} \,  \phi_{\mu\nu} \, \phi_{\rho\sigma}  \,  \phi_{\alpha \beta}
\right) \,,
\eea
where the tensors $\tilde C_\2^{\mu\nu\rho\sigma} $ and $\tilde C_\3^{\mu\nu\rho\sigma\alpha\beta} $ are defined  with the new functions 
\bea
&& \tilde a_1 = a_1 - 2 f_{2X} \, , \qquad  \tilde a_2 = a_2 + 2 f_{2X} \, ,  \\
&& \tilde b_1 = b_1 - \frac{1}{3} f_{3X} \, ,\qquad \tilde b_2 = b_2 + f_{3X} \, ,\qquad \tilde b_3 =
b_3 - \frac{2}{3} f_{3X} \,,
\eea
while  all the other functions remain unchanged. 

For the ADM decomposition of the Horndeski Lagrangians we make use of the results of Ref.~\cite{Gleyzes:2013ooa}. In particular, $L_4^{\rm H}$ and $L_5^{\rm H}$ can be rewritten, respectively, as 
\be
L_4^{\rm H} = f_2 R - (f_2 + 2 \As^2 f_{2X}) \left( K^2 - K^i_j K^j_i  \right) - 2 \As f_{2 \phi} K \;,
\ee
and
\be
\begin{split}
L_5^{\rm H} = & - \As F_3  \left( K^{ij} R_{ij} - \frac12 K R \right) - \frac13 \As^3 f_{3X} \left( K^3 - 3  K K^i_j K^j_i  + 2 K^i_j K^j_k K^k_i \right) \\
& - \frac12 \As^2 (f_{3\phi} - F_{3\phi}) R - \frac12 \As^2 f_{3\phi} \left( K^2 - K^i_j K^j_i  \right) \;, \qquad f_{3X} \equiv F_{3X} + \frac{F_3}{2 X} \;.
\end{split}
\ee

In the unitary gauge, we can 
%DL
%then
 use the relations
\beq
\A =\frac{\dot\phi}{N}\,, \qquad \V\equiv \frac1N{\cal D}_t \A\,, \qquad 
\partial_i\A=-\A \acc_i\,, \qquad \acc_i\equiv \frac{\partial_i N}{N}\,.
\eeq
%DL
%With these 
Combining all previous
 results, the ADM decomposition of the full Lagrangian in the unitary gauge leads to the following expression:
\be 
\label{LL1}
\begin{split}
{\cal L} = 
&
-\As  ^3( b_1 K^3 +b_2 K K^i_j K^j_i  + b_3 K^i_j K^j_k K^k_i) 
\\
&+ \As ^2\left[(-3 b_1+b_4 \As  ^2+f_{3X}) K^2+(-b_2+b_6 \As  ^2-f_{3X}) K^i_j K^j_i  \right]  \Vs
\\
&-\As   \left[3b_1+b_2-\As  ^2(2b_4+b_5)+\As  ^4 b_9 \right]K \Vs^2 
\\
&-  \left[b_1+b_2+b_3-\As  ^2 (b_4+b_5+b_6+b_7)+\As  ^4(b_8+b_9)-b_{10}\As   ^6\right] \Vs^3
\\
&+ \left(f_2 +a_1 \As  ^2 +\As  ^2\frac{f_{3\phi}}{2}\right) K_{ij} K^{ij} - \left(f_2 -a_2 \As  ^2 + \As  ^2\frac{f_{3\phi}}{2}\right) K^2  \\
& +\As   (4 f_{2X} +2 a_2 -a_3 \As  ^2 )  K \Vs  + \left[a_1 +\alpha_2 -(a_3 + a_4) \As  ^2 + a_5 \As  ^4\right] \Vs^2 \\
&- 2 f_{2\phi} \As  \,K - \As   F_3 \left( K^{ij} R_{ij} - \frac12 K R \right)\\
&+  \As   ^3  (3b_3-b_7 \As   ^2 - 2 f_{3X})  K^{ij} \acc_i \acc_j+ \As   ^3(2 b_2-b_5\As  ^2 + 2 f_{3,X} )K \acc_i\,\acc^i
\\
&+\As ^2  \left[2b_2+3b_3-\As  ^2(b_5+2b_6+2b_7)+\As  ^4 b_8 \right] \Vs \acc_i \,\acc^i 
\\
& +\left[ f_2-\frac{1}{2} \As  ^2 (f_{3\phi}  - F_{3\phi}) \right] R +\left[  4 f_{2X} \As  ^2 -(2 a_1 -a_4 \As  ^2)\As  ^2 \right] \acc_i \,\acc^i
\; .
\end{split}
\ee
One can then expand this Lagrangian at quadratic order around a FLRW background and  obtain an expression of the form (\ref{SBAction0}). One finds that the effective coefficients of the quadratic action are given by
\be
\begin{split}
{M^2}=&\ 2 f_2 - X[ 2 a_1+f_{3 \phi} - 6 (b_2+b_3)H\mu^2 ] \,, \\ 
\alphaT= & -1 +\frac{2f_2+X f_{3\phi}}{M^2}\;,
%\ \frac{2 X}{M^2} \left[  a_1+ f_{3 \phi} -3(b_2+b_3)H\mu^2 \right]\;,
\\
\alphaH=&-1 +\frac{1}{M^2}  \left[ 2f_2-X\left(4  f_{2X}+ 2f_{3X}H\mu^2 +f_{3\phi}\right)\right]\,, \\  
%- \frac{2 X}{M^2}  \left\{ 2  f_{2X}- a_1+[f_{3X}+3(b_2+b_3)]H\mu^2 \right\} \,, \\  
\aL   = & \ \frac{3 X}{M^2} \left[   a_1+ a_2  - (9b_1+5 b_2+ 3 b_3)H\mu^2  \right] \;, \\
\bun=& \  \frac{X}{2 M^2} \left\{ 4  f_{2X}+2 a_2 +X a_3 +2  \left[2f_{3X}-9b_1-b_2 -X (3b_4+b_6)\right] H\mu^2 \right\} \,,
\\
\bdeux=&  -  \frac{2 X}{M^2}  \left\{ a_1+a_2 + X(a_3+a_4)+ X^2 a_5 
- 3 \left[3b_1+b_2 + X(2b_4+b_5)+ X^2  b_9)\right] H\mu^2 \right\} \,,\\
\btrois=& - \frac{2 X}{M^2}  \left\{ 4  f_{2X}-2a_1- X a_4 + \left[4 f_{3X}+6b_2+3b_3+ X (3b_5+b_7)\right] H\mu^2 \right\}\, .
\end{split}
\ee
When the cubic terms are absent, these relations are equivalent to the expressions (\ref{effective}).

Using the terminology of \cite{Achour:2016rkg} and \cite{BenAchour:2016fzp}, all the subclasses of DHOST theories are summarized in Table \ref{table_DHOST}. For each subclass, we indicate the number of free functions and the degeneracy conditions verified by the effective parameters. 

\section{Degeneracy conditions}
\label{app_degeneracy}
%%%
\begin{table}
\centering
\begin{tabular}{ | c || c | c | c | }
 		\hline
 		 Subclass (see  \cite{BenAchour:2016fzp}) & $\#$ free functions & Degeneracy & Remarks\\
 		 \hline \hline
 		 \, {$^2$N-I}/Ia \, $$ & 3  &  I & H, bH \& conf-disf transf \\
 		 \, { $^2$N-II}/Ib \, $$ & 3  & 0 & \\
 		 \, { $^2$N-III}/IIa \, $$ & 3  & II & \\
 		 \, { $^2$N-IV}/IIb \, $$ & 3  &  0 & \\
 		 \hline \hline
 		 \, { $^2$M-I}/IIIa \, $$ & 3  & II & $\aT=\aH=-1$ \\
 		 \, {$^2$M-II}/IIIb \, $$ & 3  &  II & $\aL=\aT=\aH=-1$\\
 		 \, { $^2$M-III}/IIIc \, $$ & 4  & 0 &\\
 		 \hline\hline
 		\, {$^3$N-I} \, $$ & 3 & I & H, bH \& conf-disf transf  \\
 		\, { $^3$N-II} \, $$ & 4 & 0 &\\
 		\hline \hline
 		\, { $^3$M-I} \, $$ & 4  &  II& $\aT=\aH=-1$\\
 		\,  { $^3$M-II} \, $$ &   3  & II & $\aT=\aH=-1$ \\
 		\, { $^3$M-III} \, $$ &  1 & 0 & \\
 		\, { $^3$M-IV} \, $$ & 5 &  0 & \\
 		\, { $^3$M-V} \, $$ & 2 &  0 & \\
 		\, { $^3$M-VI} \, $$ & 6 &  0 & \\
 		\, { $^3$M-VII} \, $$ & 4 & 0 & \\ 
 		\hline \hline
 		\, {$^2$N-I}/Ia  \&  {$^3$N-I}\, $$ & 4  &   I  & H, bH \& conf-disf transf  \\
 		 \, {$^2$N-II}/Ib \&  $^3$N-II\, $$ & 7  & 0 & \\
 		 \hline
 		\, {$^2$N-I}/Ia  \&  {$^3$M-I}\, $$ & 5  & II & \\
 		\, {$^2$N-I}/Ia  \&  {$^3$M-II}\, $$ & 4  &  II & \\
 		\, {$^2$N-I}/Ia  \&  {$^3$M-III}\, $$ & 3  &   I \& II &  \\
 		\, {$^2$N-I}/Ia  \&  {$^3$M-V}\, $$ & 3  &  II & \\
 		\, {$^2$N-II}/Ib \&  $^3$M-VII\, $$ & 6  & 0 &\\
 		\, {$^2$N-III}/IIa \&  {$^3$M-I} \, $$ & 6  & II & \\
 		\, {$^2$N-III}/IIa \&  {$^3$M-II} \, $$ & 5  &  II & \\
 		\, {$^2$N-III}/IIa \&  {$^3$M-III} \, $$ & 4  &  II &\\
 		\, {$^2$N-III}/IIa \&  {$^3$M-V} \, $$ & 4 & II  & \\
 		\, {$^2$N-IV}/IIb \&  {$^3$M-I} \, $$ & 5  & II &\\
               \, {$^2$N-IV}/IIb \&  {$^3$M-II} \, $$ & 4  & II & \\ 
               \, {$^2$N-IV}/IIb \&  {$^3$M-III} \, $$ & 3  & 0 & \\ 
               \, {$^2$N-IV}/IIb \&  {$^3$M-V} \, $$ & 4  &  0 & \\ 
               \hline
               \, {$^2$M-I}/IIIa  \&  {$^3$M-I}\, $$ & 6  & II & $\aT=\aH=-1$\\
               \, {$^2$M-I}/IIIa  \&  {$^3$M-II}\, $$ & 5  &  II & $\aT=\aH=-1$ \\
               \, {$^2$M-I}/IIIa  \&  {$^3$M-III}\, $$ & 4  &  II & $\aT=\aH=-1$ \\
               \, {$^2$M-I}/IIIa  \&  {$^3$M-V}\, $$ & 4  &  II & $\aT=\aH=-1$ \\
 		 \, {$^2$M-II}/IIIb \&  {$^3$M-III}\, $$ & 4   &  II & $\aL=\aT=\aH=-1$  \\
 		 \, {$^2$M-II}/IIIb \&  {$^3$M-IV}\, $$ & 8   &  II & $\aL=\aT=\aH=-1$ \\
 		 \, {$^2$M-II}/IIIb \&  {$^3$M-VII}\, $$ & 6  &  II & $\aL=\aT=\aH=-1$ \\
 		 \, {$^2$M-III}/IIIc \&  {$^3$M-V}\, $$ & 6  &  0 & \\
 		 \, {$^2$M-III}/IIIc \&  {$^3$M-VI}\, $$ & 10  & 0 & \\
 		\, {$^2$M-III}/IIIc \&  {$^3$M-VII}\, $$ & 7  &  0 &\\
 		\hline
 	\end{tabular}
 \caption{Subclasses of DHOST theories, using the classification of Ref.~\cite{BenAchour:2016fzp}.  Second column: number of free functions among $f_2$, $\aq_A$, $f_3$ and $b_A$. In the degeneracy column,  $0$ stands for $M^2=0$, i.e. there are no tensor modes. If $f_{3\phi}\neq 0$, only {$^3$N-I} and {$^2$N-I}/Ia  \&  {$^3$N-I} are degenerate.}
 \label{table_DHOST}
 \end{table}

In this appendix, we concentrate on the full nonlinear action of quadratic DHOST theories. 
Their kinetic Lagrangian can be written in the form \cite{Langlois:2015cwa}
\bea\label{general kin}
L_{\rm kin} = {\cal A}(\phi,X,\A) \dot{\A}^2 + 2 {\cal B}^{ij}(\phi,X,\A) \dot{\A} K_{ij} + {\cal K}^{ijkl}(\phi,X,\A) K_{ij} K_{kl}\,,
\eea
where $ {\cal A}(\phi,X,\A) $ is a polynomial in the variable $\A$ and 
\bea
\begin{split}
 {\cal B}^{ij}(\phi,X,\A) \, &= \, {\cal B}_1 h^{ij} + {\cal B}_2 \, D^i\phi\,  D^j\phi \, , \label{calB}
 \\
 {\cal K}^{ijkl}(\phi,X,\A) \, &= \,  \kappa_1 h^{i(k} h^{j)l} + \kappa_2 \h^{ij} h^{kl} +\hat{\cal K}^{ijkl} \,.\label{calK}
 \end{split}
\eea
Here ${\cal B}_1$, ${\cal B}_2$, $\kappa_1$, $\kappa_2$ are polynomials in $\A$, and the tensor $\hat{\cal K}^{ijkl}$ vanishes when $\partial_i\phi=0$.  

The Lagrangian  is degenerate when the determinant of the kinetic matrix, 
\bea
\label{det}
D(\phi,X,\A^2) \equiv {\cal A} - {\cal K}^{-1}_{ijkl}{\cal B}^{ij} {\cal B}^{kl} = D_0(\phi,X) + \A^2 D_1(\phi,X) + \A^4 D_2(\phi,X)  \, ,
\eea
vanishes for any value of $\As$.
This  gives the three independent relations 
\bea\label{general deg}
D_0(\phi,X)=0\,, \qquad D_1(\phi,X)=0\,, \qquad D_2(\phi,X)=0 \, ,
\eea
The expressions for  $D_0$, $D_1$ and $D_2$ depend on  the functions $f_2$ and $a_A$, $(A=1,...,5)$ and on  $X$, and can be found  in eqs.~(4.30)--(4.32) of \cite{Langlois:2015cwa}.

\subsection{Degeneracy conditions in terms of the effective parameters}
For quadratic theories, one can express  $f_2$ and $a_i$ in terms of $\alphaT$, $\alphaH$, $\aL$, $\bun$,  $\bdeux$  and $\btrois$:
\be
\begin{split}
f_2 & = \frac{M^2}{2} \left(1 + \alphaT \right)\;, \qquad f_{2X}= \frac{M^2}{4X} (\alphaT- \alphaH)\;, \\ 
a_1 &= \frac{M^2}{2 X} \alphaT \;, \qquad a_2= -\frac{M^2}{6 X} (3  \alphaT - 2 \aL )\;, \\ 
a_3 &= - \frac{M^2}{3 X^2} (2 \aL -3  \alphaH -6 \beta _2 ) \;, \qquad a_4=  - \frac{M^2}{2X^2} (  2  \alphaH- \beta _3  )\;, \\ 
a_5 & = - \frac{M^2}{6X^3} (-2 \aL +3 \beta _1 +12 \beta _2 +3 \beta _3)\,.
\end{split}
\ee

Substituting these expressions into (\ref{general deg}), one gets
\begin{align}
\frac{X}{M^6} D_0 =  &  \ \frac{\aL}{3} \left[2(1+ \alphaH)^2 -(1+\alphaT)\btrois \right] \;, 
\label{D0}
\\
\frac{X^2}{M^6} D_1  = &\  \frac{\aL}{3} \big[ 4(1+\alphaH)^2- (1+\alphaT)\bdeux-(5+2\alphaT) \btrois \big] \nonumber \\ 
& -  \big[4 (1+\alphaH)\bun + 2 (1+\alphaT) \bun^2 + \btrois  \big] \;, 
\label{D1}
\\
\frac{X^3}{M^6} D_2  = & \  \frac{\aL}{3} \big[2 (1+\alphaH)^2 - (4+ \alphaT) \bdeux -(4+\alphaT) \btrois \big] \nonumber \\
&-  \big[ \bdeux + 4 (1+\alphaH)\bun + 2(4+\alphaT) \bun^2 +\btrois \big] \;.
\label{D2}
\end{align}

One can distinguish two cases, which yield the conditions $\CI$ and $\CII$, given respectively in (\ref{Ia}) and (\ref{IIa}).
\begin{itemize}
\item $\aL=0$: The condition (\ref{D0}) is automatically satisfied. From (\ref{D1}) and (\ref{D2}) we obtain 
\be \label{DegCond1}
\btrois=-4\bun(1+\alphaH)-2\bun^2(1+\alphaT) \; , \qquad  \bdeux=-6\bun^2 \; .
\ee

\item $\aL \neq 0$: The three conditions can be solved for $\bun$, $\bdeux$ and $\btrois$, obtaining
\be\label{DegCond2}
\begin{split}
({1+\alphaT})\bun & =-(1+\aL) ({1+\alphaH})  \; , \qquad \left({1+\alphaT}\right)^2 \bdeux=-6(1+\aL) \left({1+\alphaH} \right)^2   \; , \\
({1+\alphaT}) \btrois & = 2 (1+\alphaH)^2 \;.
\end{split}
\ee
\end{itemize}

\subsection{Unitary gauge and extra degree of freedom}
\label{app_degeneracy_2}
In the unitary gauge, we have $\partial_i\phi=0$ and therefore $X=- \A^2$ so that the vanishing of the determinant (\ref{det}) is guaranteed by the single condition
\bea
\label{uni_deg}
D(\phi,X,-X)=0 \,. 
\eea
This condition is obviously satisfied if the three conditions of \eqref{general deg} are verified, but the converse is not true: there exist theories that look degenerate in the unitary gauge but are in fact not degenerate.

An important consequence of the above discussion is that for a nondegenerate theory that satisfies  (\ref{uni_deg}), the extra scalar degree of freedom, which is known to be present because the theory is nondegenerate, does not show up in the linear perturbations about a homogeneous background. Indeed, in this case  the kinetic Lagrangian quadratic in perturbations is given by 
\bea
L_{\rm kin}^{\rm quad} = \bar{\cal A} \, \delta \dot{\A}^2 + 2 \bar{\cal B}^{ij}\, \delta \dot{\A} \delta K_{ij} + \bar{\cal K}^{ijkl} \, \delta K_{ij} \delta K_{kl}\,,
\eea
where all quantities are decomposed into a background component, denoted by a bar,  and a perturbative component: $\phi=\overline{\phi} +\delta \phi$, $X=\overline{X} + \delta X$ and $\A=\overline{\A} + \delta \A$. Note that $\delta X$ and $\delta \A$ involve $\delta \dot \phi$
and $\partial_i \delta \phi$. Since $\bar X=-\bar\A^2$, we see that the degeneracy of the above quadratic Lagrangian is automatically guaranteed when the condition (\ref{uni_deg}) is verified, even if the conditions (\ref{general deg}) are not. 
The condition can be written as 
\be
\bar{\cal A}-\frac{3}{\bar\kappa_1+3\bar\kappa_2}\bar{\cal B}_1^2=0\,. 
\ee
In that case, it means that only one scalar degree of freedom shows up at the level of linear perturbations, independently of the gauge choice. However, the extra degree of freedom remains present and would show up at the nonlinear level or in an inhomogeneous background.

\section{Disformal transformations}
\label{app:disformal}
\newcommand{\tilbN}{\tilde{N}_{\rm b}}
\newcommand{\fN}{\Xi}
\newcommand{\delXtil}{ \frac{\delta \tilde{N}}{\tilde{N}} }
\newcommand{\delXtilP}{{\left(\frac{\delta \tilde{N}}{\tilde{N}_{b}}\right)}^{\dot{}}}
\newcommand{\delXtilPsq}{{\left(\frac{\delta \tilde{N}}{\tilde{N}_{b}}\right)}^{\dot{}\,2}}
\newcommand{\parDelXtil}{\left(\frac{\partial_i \delta \tilde{N}}{\tilde{N}_{b}}\right)}
\newcommand\FouralphaYY{\frac{2}{C} \frac{\partial^2 C}{\partial N ^2}}

Here we study how action \eqref{SBAction0} transforms under a conformal and disformal redefinition of the metric, eq.~(\ref{disf_cov}).
For simplicity, following \cite{Gleyzes:2014qga} we work in the unitary gauge and set $\phi = t$ (i.e.~$\mu=1$ in eq.~\eqref{muphi}), so that $X = - 1/N^2$. 
In the unitary gauge, the disformal  transformation (\ref{disf_cov}) is of the form
\be
\tilde{g}_{\mu \nu} = C(t, N) g_{\mu \nu}  + D(t , N) \delta_\mu^0 \delta_\nu^0 \;.
\ee
The ADM components of the  metric $\tilde{g}_{\mu \nu}$ are related to those of $g _{\mu \nu}$ by
\be \label{disformalADM}
\tilde{N^i}=N^i \, ,\qquad \tilde{h}_{ij}=C h_{ij} \, , \qquad  {\tilde{N}}^2=C N^2-D \, .
\ee
Moreover, the other relevant geometrical quantities  in the action transform as 
\begin{align}
\sqrt{-\tilde{g}} &= \sqrt{-g} \, C^{3/2} \sqrt{C-D/N^2} \; , \\
\tilde {R}&=C^{-1} \left( R -2D^i D_i\log{C}-\frac{1}{2}(\partial_i \log{C})^2\right)\; , \\
\tilde{K}^i_j&=\frac{N}{\tilde{N}} \left[ K^i_j +\frac{1}{ 2 N C} \left(\dot{C}+C_N(\dot{N}-N^i\partial_iN)\right)\delta^i_j \right]\; . \label{disformalRK} 
\end{align}

The dimensionless time-dependent parameters defined in eq.~\eqref{defalphas2} become
\be
\begin{split}
&\alphaC \equiv \frac{\dot C}{2 H C N} \; , \qquad \alphaY \equiv \frac{N}{2 C} \frac{\partial C}{\partial N} \, , \qquad \alphaD \equiv \frac{D}{ CN^2-D}\, , \qquad \alphaX \equiv - \frac{1}{2 C N} \frac{\partial D}{\partial N}\; ,
 \end{split}
\ee
where all the quantities on the right-hand side are evaluated on the background. 
Using these definitions, from eq.~\eqref{disformalADM} the homogeneous components of the metric and the lapse transform as
\be\label{Nbtil}
\tilde a = \sqrt{C} a \;, \qquad \tilde N_0 =\sqrt{\frac{C}{1+\alphaD}} N_0 \; .
\ee
Moreover, the Hubble rate changes accordingly, i.e.,
\be \label{Hbar}
\tilde{H}\equiv \frac{1}{\tilde N_0 \tilde{a}} \frac{d \tilde{a}}{d t}  =H (1+\alphaC)\sqrt{\frac{1+\alphaD}{C}} \; .
\ee
Furthermore, to compute how the action transforms we are interested in the following relations, derived from linearizing eqs.~\eqref{disformalADM}--\eqref{disformalRK} and valid at linear order,
\be
\begin{split}
\label{lineartrans}
\delta N=& \  \fN  \frac{N}{\tilde N} \delta \tilde N \; ,\\
\delta \sqrt{h}=& \ \frac{a^3 }{\tilde{a}^3} \delta \sqrt{\tilde{h}} -3 a^3 \fN  \alphaY  \delXtil \; ,\\
R= & \ C \tilde{R}+4  \alphaY   \frac{\fN}{  a^2}  \frac{\Delta \delta \tilde N}{\tilde N} \; , \\
\delta K^i_j=& \ \frac{\tilde N}{N} \,\delta \tilde{K}^i_j-  \alphaY\,\frac{\fN }{N}\, \frac{d}{dt} \bigg( \delXtil \bigg) \delta^i_j+ \delta^i_j \,H \Upsilon \,\delXtil\; , 
\end{split}
\ee
where
\be
\fN \equiv  \ \frac{1}{(1+\alphaD)(1+\alphaX+\alphaY)} \; ,
\ee
and
\be
\begin{split}
\Upsilon \equiv & \ \fN \bigg\{ \alphaX+\alphaY+\alphaD(1+\alphaX+\alphaY)+\alphaC\left[(1+\alphaD)(1+\alphaX)+\alphaY (3+\alphaD) \right] \\
&-\frac{1}{2 H C} \frac{\partial^2 C}{\partial t \partial N} -\alphaY\frac{\dot{\fN} }{ \fN H } +\frac{\dot \bN}{H \bN } \left[\alphaX\alphaY+3\alphaY^2+\alphaD\alphaY(1+\alphaX+\alphaY)-\frac{\bN^2}{2 C} \frac{\partial^2 C}{\partial N^2}\right] \bigg\} \; .
\end{split}
\ee
The last ingredient is the transformation of the second-order perturbation of $R$, i.e.
\be
\begin{split}
\label{R2trans}
\delta_2 R=& \ C  \delta_2 \tilde{R}+2 C  \alphaY  \fN \tilde{R}  \delXtil- 4\fN \alphaY \frac{\delta \sqrt{\tilde{h}}}{\tilde{a}^3} \frac{ \Delta \delta \tilde N}{ a^2 \tilde N} +\left( \FouralphaYY-18 \fN^2  \alphaY^2\right)  \frac{(\partial_i \delta \tilde N)^2}{ a^2  \tilde N^2}\\
&+\frac{\delta\tilde N\,  \Delta \delta\tilde N}{ a^2  \tilde N^2} \left( \FouralphaYY-8 \alphaY^2\right) \;.
\end{split}
\ee

Replacing all the geometrical terms in action \eqref{SBAction0} by their transformed quantities using eqs.~\eqref{lineartrans} and \eqref{R2trans} and making use of eqs.~\eqref{Nbtil} and \eqref{Hbar} for the homogeneous quantities, one obtains an action in terms of the metric $\tilde g_{\mu \nu}$.
The last step is to make a time redefinition, 
\be
t \to \tilde t = \int  \sqrt{\frac{C}{1+\alphaD}} dt \;,
\ee
which sets the homogeneous ``00'' component of the metric to unity, i.e.~$\tilde N_0 = 1$.
The obtained action $\tilde S$ has the same form as the original one, eq.~\eqref{SBAction0}, with the transformed $\tilde \alpha_{\rm K}$ and $\tilde \alpha_{\rm B}$ given by 
\be 
\begin{split}
\tilde{\alpha}_{\rm K} &=\frac{ \alphaK \fN^2+12 \fN  \Upsilon  \alphaB-6 \Upsilon^2 (1+\aL)}{ (1+\alphaC)^2}+12 \Upsilon  \bun  \frac{ \dot{\fN}}{H  (1+\alphaC)^2} +\bdeux \frac{\dot{\fN}^2 }{H^2 (1+\alphaC)^2}\\
&-\frac{1}{2 \tilde{M}^2 \tilde{a}^3 \tilde{H}^2} \frac{d}{d t} \bigg\{ \tilde{M}^2 \tilde{a}^3 \frac{12 \fN H  }{\tilde N_0}\bigg[\Upsilon (\alphaY (1+\aL) +\bun )-\fN \alphaB \alphaY+\frac{\dot{\fN}}{6 H }  \big(\bdeux -6 \alphaY \bun \big) \bigg] \bigg\} \;, \\ 
\tilde{\alpha}_{\rm B} &= \frac{ \alphaB \fN  - (1+\aL)  \Upsilon}{1+\alphaC} +\frac{\bun\, \dot{\fN}}{H (1+\alphaC)} \;.  
\end{split}
\ee
The remaining time-dependent functions $\tilde \alpha_A$ and $\tilde \beta_A$ are given by eq.~\eqref{alphatilde}.

%%%%%%%%%%%%%%%%%%%%%%
\section{Quadratic Lagrangian and dispersion relation in the presence of matter}
\label{app:matter}
\label{app:elements}

\newcommand\alphaSm{\alpha^{\rm eff}_{\text{D,m}}}
\newcommand\alphaDm{\alpha_{\text{D,m}}}
\newcommand\alphaCm{\alpha_{\text{C,m}}}
\newcommand\alphaXm{\alpha_{\text{X,m}}}
\newcommand\alphaYm{\alpha_{\text{Y,m}}}
\newcommand\alphaS{\alpha^{\rm eff}_{\text{D}}}

We describe matter using a scalar field $\sigma$ with a $k$-essence type action \cite{ArmendarizPicon:2000dh},
\be
S_{\rm m} = \int d^4 x \sqrt{-g} P(Y) \;,   \qquad Y \equiv \tg^{\mu \nu} \partial_\mu \sigma \partial_\nu \sigma \;,
\ee
where matter is coupled to a metric of the form
\be
\label{disf_unit_I2_m}
\tg_{\mu \nu} = C_{\rm m}^{(\phi)}(\phi, X) g_{\mu \nu}  + D^{(\phi)}_{\rm m}(\phi, X) \partial_\mu \phi \, \partial_\nu \phi \;.
\ee
In unitary gauge, this reads
\be
\label{disf_unit_I_m}
\tg_{\mu \nu} = C_{\rm m}(t, N) g_{\mu \nu}  + D_{\rm m}(t , N) \delta_\mu^0 \delta_\nu^0 \;,
\ee
with
\be
C_{\rm m}(t,N) =  C_{\rm m}^{(\phi)} \big( \phi( t)  ,-\dot \phi( t)^2/N^2\big)\, , \qquad D_{\rm m}(t, N) =   \dot{ \phi}^2 (t) D_{\rm m}^{(\phi)} \big( \phi( t) , -\dot \phi( t)^2/N^2 \big)\,.
\ee
Then, in analogy with equation~\eqref{defalphas2}, we introduce the parameters
\be
\label{defalphas_m}
 \alphaCm \equiv \frac{\dot C_{\rm m} }{2 H  C_{\rm m}} \, ,\quad \alphaYm \equiv \frac{1}{2 C_{\rm m}} \frac{\partial C_{\rm m} }{\partial N}\, , \quad \alphaXm \equiv - \frac{1}{2  C_{\rm m} } \frac{\partial D_{\rm m} }{\partial N} \,  , \quad  \alphaDm \equiv \frac{D_{\rm m}}{  C_{\rm m}-D_{\rm m}}\,,
 \ee
where the right-hand sides are evaluated on the background.
The first two parameters in the above equations, $\alphaCm$ and $\alphaDm$, were introduced in Ref.~\cite{Gleyzes:2015pma}, while the third in~\cite{DAmico:2016ntq}.
%The matter energy density and pressure are given by
%\be
% \rho_{\rm m} = C_{\rm m}^2\sqrt{1+\alphaDm}\left(2 Y P_Y - P\right)\,, \qquad P_{\rm m}=\frac{C_{\rm m}^2\,P}{\sqrt{1+\alphaDm}}\,,
%\ee
The total action is the sum of the gravitational action, eq.~\eqref{SBAction0}, and the matter action above. 
We will first discuss the case $\aL \neq 0$; the case $\aL =0$ can  be obtained by taking the smooth limit $\aL \to 0$, as we will discuss below.

Variation of the total action with respect to $\psi$ yields the scalar component of the momentum constraint, which reads
\be\label{psiconstr}
\begin{split}
&2 \aL \frac{\Delta \psi}{a^2} + 6 H \delta N ( 1 + \alphaB+ \aL) +6 \bun {\delta \dot N}- 6 (1+\aL) \dot{\zeta}=3(1+\alphaDm)c_{\rm m}^2 \dsz {\Az}_{  \rm m}\, \dels
%-6 \frac{C_{\rm m} \sqrt{1+\alphaDm} \dsz P_{Y}}{M^2}\, \dels\; ,
\end{split}
\ee
where $\sigma_0=\sigma_0(t)$ and $\delta \sigma=\delta \sigma(t,\bold{x}) $ respectively denote the homogeneous component of the scalar field and its perturbation. We also introduced the matter equation of state and sound speed squared including a lightcone-changing factor $ \left(1+\alphaDm\right)$ for later convenience,
\be
w_{\rm m} =\frac{P}{\left(1+\alphaDm\right)(2  Y P_Y -P )} \;, \qquad c_{\rm m}^2 = \frac{P_Y}{ \left(1+\alphaDm\right)\left(P_Y + 2 Y P_{YY} \right)} \;,
\ee
as well as the combination
\be\label{Amdef}
{\Az}_{  \rm m}\equiv  -\frac{2 C_{\rm m} \sqrt{1+\alphaDm}}{M^2 } \left(P_Y + 2 Y P_{YY} \right) \;.
\ee
%\be\label{Amdef}
%{\Az}_{  \rm m}\equiv  \frac{3  \big[1+w_{\rm m}(1+\alphaDm)\big] \Omega_{\rm m} H^2}{  c_{\rm m}^2 \dsz^2 (1+\alphaDm)^2}\;
%\ee
%and we have defined the fractional matter energy density,
%\be
%\Omega_{\rm m}  \equiv \frac{\rho_{\rm m}}{3 H^2 M^2 } \;.
%\ee
Substituting  $\psi$ into the total  action using equation~\ref{psiconstr} yields
\be\label{actionConstrPsiM}
\begin{split}
S =\int & d^3x \, dt \,a^3 \frac{M^2}{2}    \bigg\{ \frac{1}{  \aL }\bigg[ 6 (1+\aL)  \dot \zeta ^2  -1 2 \bun   \delta \dot N  \dot \zeta + (6 \bun^2 + \aL \bdeux )  \delta \dot N ^2  \\
& +12  H   \left(\bun   \left(1 + \alphaB \right)  \delta \dot N - \left(1 + \alphaB + \aL \right) \dot \zeta \right)  \delta  N  \\
& + H^2  \left( 6  \alphaB^2+12  \alphaB +  \aL \alphaK+6 (1+\aL) +\frac{\aL \MMM_{22} ^{(\rm m)}}{H^2}\right)  \delta  N ^2\bigg] \\
&  +\frac{1}{a^2}  \bigg[ 4 \left(1+ \alphaH\right) \partial_i \zeta \partial_i \delta  N   + 2 \left(1+  \alphaT \right) \left(\partial_i  \zeta \right)^2 +\btrois  \left(\partial_i  \delta N\right)^2  \bigg] \\
&+ {\Az}_{  \rm m}\, \bigg[ \delta \dot \sigma^2 - \frac{c_{\rm m}^2}{a^2}  \left(\partial_i  \dels \right)^2+6\dsz  c_{\rm m}^2 (1+\alphaDm) \left( \zeta \, \delta \dot \sigma +\frac{1+\aL}{\aL} \dot{\zeta}\; \dels \right)\\
&-2 \dsz (1+\alphaDm) \left( \delta N \;\delta \dot \sigma \big(1+\alphaXm+\alphaYm(1-3c_{\rm m}^2)\big)+\frac{3 c_{\rm m}^2 \bun}{\aL} {\delta \dot N}\;\dels \right)\\
&-\frac{6 H (1+ \alphaB+\aL)\dsz c_{\rm m}^2  (1+\alphaDm)}{ \aL}\delta N \dels+
3\frac{c_{\rm m}^4 \dsz^2 (1+\alphaDm)^2}{2 \aL}{\Az}_{  \rm m}\dels^2 \bigg] \bigg\}\, ,
\end{split}
\ee
where we have defined 

\be
\begin{split}
\MMM_{22} ^{(\rm m)}=&\ 3 \hat{\Omega}_{\rm m} H^2 \alphaSm+
 \dsz^2 (1+\alphaDm)^2{\Az}_{  \rm m}\left[(1+\alphaXm+\alphaYm) -3c_{\rm m}^2(\bun+\alphaYm)\right]^2\; , \\
 \hat{\Omega}_{\rm m} =&\ \frac{C_{\rm m}^2\sqrt{1+\alphaDm}\left(2 Y P_Y - P\right)}{3 M^2 H^2}\, ,\\
\alphaSm\equiv&\ \alphaDm(1+\alphaXm+\alphaYm)^2 +2(\alphaXm -\alphaYm)+(\alphaXm+\alphaYm)^2+\frac{1}{2 C_{\rm m}}\frac{\partial^2 (D_{\rm m}-C_{\rm m})}{\partial N^2} \\
&+ \big[1+w_{\rm m}(1+\alphaDm)\big]\Big[6 \bun(1+\alphaXm+\alphaYm) -9c_{\rm m }^2(\bun+\alphaYm)^2\Big]\\
&+3 w_{\rm m} \Big[2\alphaYm (1+\alphaDm)(1+\alphaXm)+\alphaYm^2(3+2 \alphaDm)\\
&+2\bun(1+\alphaDm )(1+\alphaXm+\alphaYm)+\frac{1}{2 C_{\rm m}}\frac{\partial^2 C_{\rm m}}{  \partial N^2} \Big]
\; .
\end{split}
\ee

Let us now impose the unitary degeneracy condition, eq.~\eqref{uni_degeneracy}. As discussed in Sec.~\ref{sec:cosmo}, it is convenient to introduce $\tzeta$ defined in eq.~\eqref{tzetadef}. Replacing $\zeta$ in terms of this variable in action \eqref{actionConstrPsiM} eliminates time derivatives of the lapse $\delta N$. In particular, the quadratic action takes the form
\be 
\label{SxyxM}
\begin{split}
S=\int  d^3x \, dt \, a^3 \frac{M^2}{2}\,   \bigg[ & \frac{6 (1+\aL)}{\aL} \dot{\tzeta}^2+ {\Az}_{  \rm m} \delsD^2  + 2 \dot{\mathbf{ X}}^T \VVV \mathbf{ X}+ \frac{1}{a^2} \partial_i \mathbf{ X}^T  \SSS \, \partial_i \mathbf{ X} +   \mathbf{ X}^T \MMM \,  \mathbf{ X}   \bigg] \; , 
\end{split}
\ee
where  $\mathbf{X}^{T} \equiv(\tzeta,\delta N, \dels)$ and ${\cal V}$ denotes the matrix of elements
\be
\begin{split}
\VVV_{12} &= -\frac{6} {\aL}\left[H  (1 + \alphaB+ \aL)- \dot \bun+\frac{ \bun \dot {\alpha}_{\rm L} }{1+\aL}\right]  \; , \qquad
\VVV_{13}  = \frac{ 3 (1+\aL)}{\aL}\dsz  (1+\alphaDm)c_{\rm m}^2 {\Az}_{  \rm m}\; , \\
\VVV_{32} &=  - \dsz (1+\alphaDm){\Az}_{  \rm m}
 \left[1+\alphaXm+\alphaYm- 3c_{\rm m}^2\left( \frac{  \bun}{1+\aL}+\alphaYm\right) \right]\; , \\
\VVV_{11}  &= \VVV_{21} = \VVV_{22} =  \VVV_{23} = \VVV_{31}  = \VVV_{33} =0 \;.
\end{split}
\ee 
Moreover, $\MMM$ and $\SSS$ are symmetric matrices with elements given by
\be
\begin{split}
\SSS_{11} & =  2(1+\alphaT)\;, \qquad \SSS_{12}    =  2 \left[1+\alphaH+\frac{(1+\alphaT) \bun}{1+\aL}\right]\;, \qquad  \SSS_{13}  =\SSS_{23}=0  \;, \\
\SSS_{22} & = 4\frac{(1+\alphaH) \bun }{1+\aL}+2\frac{(1+\alphaT) \bun^2 }{(1+\aL)^2}+\btrois \;, 
\qquad \SSS_{33}  = \,-c_{\rm m}^2 {\Az}_{  \rm m}\; ,\\
\end{split}
\ee
and 
\be
\begin{split}
\MMM_{22}  = &\ \frac{\aL\VVV_{12}}{1+\aL} \left(\frac{\VVV_{12}}{6} - H  \alphaB  \right)+H^2 \left[ \alphaK - 6  \alphaB \left(1+ \frac{3+\alphaM}{1+\aL}\bun\right)\right] 
 - 6 \frac{(\alphaB H )^{\hbox{$\cdot$}}  }{1+\aL}\bun  +\MMM_{22} ^{(\rm m)}\; ,\\
%&+w_{\rm m}(1+\alphaDm)\Big(1+\alphaXm+\alphaYm-3c_{\rm m}^2 (\bun+\alphaYm) \Big)^2\\
%&+c_{\rm m}^2\Big(\alphaSm-6(\bun+\alphaYm)(1+\alphaXm+\alphaYm)\Big)\Bigg]\; , \\
\MMM_{23}   = &\ \frac12\dsz  (1+\alphaDm)c_{\rm m}^2 {\Az}_{  \rm m} \VVV_{12}\;, \qquad \MMM_{33}   = \frac{3}{2\aL} \dsz^2  (1+\alphaDm)^2 c_{\rm m}^4 {\Az}^2_{  \rm m}\;, \\
\MMM_{11}  =&\ \MMM_{12}=\MMM_{13}=0  \;.
\end{split}
\ee
%in the second equation above, we have defined
%\be
%\begin{split}
%\alphaSm\equiv&\ \alphaDm(1+\alphaXm+\alphaYm)^2 +2(\alphaXm -\alphaYm)+(\alphaXm+\alphaYm)^2+\frac{1}{2 C_{\rm m}}\frac{\partial^2 (D_{\rm m}-C_{\rm m})}{\partial N^2} \\
%&+ \big[1+w_{\rm m}(1+\alphaDm)\big]\Big[6 \bun(1+\alphaXm+\alphaYm) -9c_{\rm m }^2(\bun+\alphaYm)^2\Big]\\
%&+3 w_{\rm m} \Big[2\alphaYm (1+\alphaDm)(1+\alphaXm)+\alphaYm^2(3+2 \alphaDm)\\
%&+2\bun(1+\alphaDm )(1+\alphaXm+\alphaYm)+\frac{1}{2 C_{\rm m}}\frac{\partial^2 C_{\rm m}}{  \partial N^2} \Big]
%\; .
%\end{split}
%\ee
Variation of~\eqref{SxyxM} with respect to $\delta N$ gives the Hamiltonian constraint,
\be \label{constr2M}
\VVV_{12} \dot \tzeta+\VVV_{23} \dot \dels-\SSS_{21} \Delta  \tzeta -\SSS_{22} \Delta  \delta N +\MMM_{22} \delta N +\MMM_{23} \dels=0 \;.
\ee
This can be solved for $\delta N$ and plugged back into  action~\eqref{SxyxM}. After a spatial Fourier transform we get, denoting $\hat{k}=k/a$:
\be \label{SxM}
\begin{split}
S=& \frac{1}{2(2\pi)^3}\int dt d^3k   \frac{a^3M^2}{\MMM_{22}+\hat{k}^2 \SSS_{22}}  \bigg[ \Big (\aone+\bone \hat{k}^2\Big)   \, \dot{\tzeta}_{\mathbf{k}} \, \dot{\tzeta}_{\mathbf{-k}} + \frac{\btwo \hat{k}^2+\ctwo \hat{k}^4+\dtwo \hat{k}^6}{\MMM_{22}+\hat{k}^2 \SSS_{22}}  \, \tzeta_{\mathbf{k}}\,\tzeta_{\mathbf{-k}} \\
&+ \Big(\athr+\bthr \hat{k}^2\Big)  \, \dot{\dels}_{\mathbf{k}} \, \dot{\dels}_{\mathbf{-k}}+ \frac{\afour+\bfour \hat{k}^2+\cfour \hat{k}^4+\dfour \hat{k}^6}{\MMM_{22}+\hat{k}^2 \SSS_{22}}  \, \dels_{\mathbf{k}}\,\dels_{\mathbf{-k}}\\
&+2 \afive \, \dot{\tzeta}_{\mathbf{k}} \, \dot{\dels}_{\mathbf{-k}}+ 2 \bsix \hat{k}^2 \, \tzeta_{\mathbf{k}} \, \dot{\dels}_{\mathbf{-k}}+2\Big(\asev +\bsev \hat{k}^2\Big) \, \dels_{\mathbf{k}} \, \dot{\tzeta}_{\mathbf{-k}}+2  \beight \hat{k}^2 \, \tzeta_{\mathbf{k}} \, \dels_{\mathbf{-k}} \bigg] \; ,
\end{split}
\ee
where
\be\label{ABCDE}
\begin{split}
\aone=&\ \frac{6 (1+\aL)}{\aL} \MMM_{22}-\VVV_{12}^2\; ,\qquad\bone= \frac{6 (1+\aL)}{\aL} \SSS_{22}  \; ,\\
\btwo=&\ \SSS_{11} \MMM_{22}^2+H \SSS_{12}\MMM_{22}\VVV_{12}(3+\alphaM) +\MMM_{22} \VVV_{12} \dot{\SSS}_{12}-\VVV_{12} \SSS_{12} \dot{\MMM}_{22}+\SSS_{12} \MMM_{22} \dot{\VVV}_{12}\; ,\\
\ctwo=&-\SSS_{12}^2 \MMM_{22}+\SSS_{22} \left(2\SSS_{12}\MMM_{22}+\VVV_{12}\dot{\SSS}_{12}\right)+\SSS_{12}\left(H\SSS_{22}\VVV_{12}(3+\alphaM)-\VVV_{12}\dot{\SSS}_{22}+\SSS_{22}\dot{\VVV}_{12}\right)\; ,\\
\dtwo=&\ \SSS_{22} \Big(\SSS_{11}\SSS_{22}-\SSS_{12}^2\Big) \; ,\qquad \athr= {\Az}_{  \rm m}\MMM_{22}-\VVV_{32}^2\; ,\qquad \bthr=  {\Az}_{  \rm m} \SSS_{22}\; ,\\
\afour=& - \MMM_{22} \left[ \MMM_{23}^2 + \big(\MMM_{33}\MMM_{22}+\VVV_{32}\dot{\MMM}_{23}\big) \right]+\MMM_{23}\left[H \MMM_{22}\VVV_{32}\big(3+\alphaM\big)-\VVV_{32}\dot{\MMM}_{22}+\MMM_{22}\dot{\VVV}_{32}\right]\; ,\\
\bfour=&\ \SSS_{33} \MMM_{22}^2-\MMM_{23}\VVV_{32}\dot{\SSS}_{22} +\SSS_{22}\left[2\MMM_{33}\MMM_{22}-\MMM_{23}^2+\VVV_{32 }\dot{\MMM}_{23}+\MMM_{23}\big(H \VVV_{32}(3+\alphaM)+\dot{\VVV}_{32} \big)\right]\; ,\\
\cfour=&\  \SSS_{22}\big(2\SSS_{33} \MMM_{22}+\SSS_{22} \MMM_{33}\big)\; ,\qquad \dfour= \SSS_{22}^2 \SSS_{33}\; ,\qquad \afive=-\VVV_{12} \VVV_{32} \; ,\\
%\asix=&- \VVV_{32} \MMM_{21}  \; ,\\
\bsix=& - \VVV_{32} \SSS_{12}  \; ,\qquad \asev= \VVV_{13}\MMM_{22}- \VVV_{12} \MMM_{23} \; ,\qquad
\bsev= \VVV_{13}\SSS_{22}\; ,\qquad
%\aeight=& \MMM_{22} \MMM_{31}-\MMM_{21}\MMM_{23}\; ,\\
\beight= - \SSS_{12}\MMM_{23}\; .\\
\end{split}
\ee

Let us focus on theories satisfying the degeneracy conditions $\CI$. By imposing the  remaining degeneracy conditions,   the limit $\aL \to 0$ is finite and for the kinetic part of the action we find
\be
\begin{split}
S= &\int d^3 x \,dt \, a^3   \frac{M^2 }{2} \bigg\{  {\Az}_{\tzeta} ^{(\rm m)}  \dot{\tzeta}^2 +   \frac{{\Bz}_{\tzeta}}{a^2}(\partial_i \tzeta)^2 + {\Az}_{  \rm m}  \left[ \delta \dot \sigma^2 - \frac{{c}^2_{  \rm m}}{a^2}  \left(\partial_i  \dels \right)^2 \right] +{\Az}_{\tzeta  \rm m} \dot \dels \dot \tzeta+\frac{{\Bz}_{\tzeta \rm m}}{a^2}\partial_i \tzeta \,\partial_i \dels \bigg\}\; ,
\end{split}
\ee
where
\begin{align}
{\Az}^{(\rm m)}_{\tzeta} & =  {\Az}_{\tzeta}+\frac{\MMM_{22} ^{(\rm m)}}{\big[H(1+\alphaB)-\dot \bun \big]^2}\; , \\
{\Az}_{\tzeta  \rm m}& = \frac{2  \dsz (1+\alphaDm) {\Az}_{  \rm m}  \big[1+\alphaXm+\alphaYm-3c_{\rm m}^2(\bun+\alphaYm)\big]}{   \big[H(1+\alphaB)-\dot \bun \big]^2} \; ,\\
{}\Bz_{\tzeta \rm m}& =- \frac{2 \dsz (1+\alphaDm) c_{\rm m}^2 {\Az}_{  \rm m}   \big[1+\alphaH+\bun(1+\alphaT)\big]}{   \big[H(1+\alphaB)-\dot \bun \big]^2}\; ,
\end{align}
and $\Az_{\tzeta}$, $\Bz_{\tzeta}$ are respectively defined in eqs.~\eqref{AzI} and \eqref{BzI}.
Requiring that the time kinetic matrix is positive definite gives the no-ghost conditions
\be
{\Az}_{  \rm m} \geq0 \; , \qquad \alpha +3\hat{\Omega}_{\rm m}\alphaSm \geq0 \; ,
\ee
where $\alpha$ was defined in equation~\eqref{alphaZ}. 
Requiring that the determinant of the kinetic matrix vanishes gives the dispertion relation 
\be
(\omega^2-\hat{c}_{\rm s}^2 k^2)(\omega^2-c_{\rm m}^2 k^2)=\lambda^2 \hat{c}_{s}^2 \omega^2 k^2 \; ,
\ee
where 
\be\label{lambda}
\lambda^2\equiv\frac{3  \big[1+w_{\rm m}(1+\alphaDm)\big] \hat{\Omega}_{\rm m} H^2 } {\alpha+3\hat{\Omega}_{\rm m}\alphaSm}\, \big[\alphaH-\alphaXm-\alphaYm+3c_{\rm m}^2(\alphaYm+\bun)+\bun(1+\alphaT)\big]^2
\ee
is a frame-independent parameter giving the amount of kinetic mixing between matter and the scalar field, which generalizes the one introduced in~\cite{DAmico:2016ntq} (and to which it reduces in the limit $\alphaYm =\bun=0$). 
Moreover, in equation~\eqref{lambda}, $\hat{c}_{s}^2$ is defined as
\be
\hat{c}_{s}^2  \equiv \frac{\alpha}{\alpha+3\hat{\Omega}_{\rm m}\alphaSm }c_{s}^2 -\frac{3  \big[1+w_{\rm m}(1+\alphaDm)\big] \hat{\Omega}_{\rm m}  } {\alpha+3\hat{\Omega}_{\rm m}\alphaSm}\, \big[1+\alphaH+\bun(1+\alphaT)\big]^2 \; ,
\ee
where $c_{s}^2 $ is the sound speed in the absence of matter, defined in eq.~\eqref{cs0Ia}. 
%{\Az}_{\tzeta} & = \frac{  1}{(1+\alphaB-\dot \bun /H)^2}  \bigg[\alphaK+ 6\alphaB^2 - \frac{6}{a^3 H^2 M^2} \frac{d}{dt} \left( a^3 H M^2 \alphaB \bun \right) \bigg] \; , \\

%%%%%%%%%%%%%%%%%%%%%%%%%%%%%%%%%%%%%%%%%%%%%%%%%%%%%%%%%%%%%%%%%%%%%%%%%

\bibliographystyle{utphys}
\bibliography{EFT_DE_biblio2}

%\begin{thebibliography}{99}

%\end{thebibliography}

\end{document}